\documentclass[aps,prd,groupedaddress,preprint,eqsecnum,nofootinbib]{revtex4}
\usepackage{graphicx,epsf,amssymb,amsbsy,amsfonts,amssymb,amsmath,physics,courier,IEEEtrantools}

\def\be{\begin{equation}}
\def\ee{\end{equation}}

\def\bg{\bar{g}}
\def\beq{\begin{eqnarray}}\def\eeq{\end{eqnarray}}
\def\ba#1\ea{\begin{align}#1\end{align}}
\def\bg#1\eg{\begin{gather}#1\end{gather}}
\def\bm#1\em{\begin{multline}#1\end{multline}}
\def\bmd#1\emd{\begin{multlined}#1\end{multlined}}

\def\({\left(}
\def\){\right)}
\def\[{\left[}
\def\]{\right]}

\pdfoutput=1

\begin{document}
\hfuzz 9pt
%\maketitle
\title{Symmetries of free massless particles and soft theorems}
\author{Shamik Banerjee}
%\date{}                                           % Activate to display a given date or no date
\affiliation{Institute of Physics, \\ Sachivalaya Marg, Bhubaneshwar, India-751005 \\ and \\ Homi Bhabha National Institute, Anushakti Nagar, Mumbai, India-400085}

\email{banerjeeshamik.phy@gmail.com}
\begin{abstract}
In an earlier paper we have constructed a basis of massless single particle quantum states which transform in the unitary principal series representation of the four dimensional Lorentz group. The S-matrix written in this basis gives rise to the Mellin transformed amplitude of Pasterski-Shao-Strominger and its generalization. In this basis the particle can be thought of as living on the null-infinity in the Minkowski space. In this paper we take some preliminary steps to see how the connection between soft theorems and symmetries work out in this picture. For simplicity we consider only the leading soft photon and soft graviton theorems which are related to U(1) Kac-Moody and supertranslations.

\end{abstract}

%\preprint{}
\maketitle
\tableofcontents

\title{Soft Theorems}
\author{Shamik Banerjee}
\date{}							% Activate to display a given date or no date

%\begin{document}
\maketitle
%\section{}
%\subsection{}
\section{Introduction}

In the pioneering works \cite{Strominger:2013lka,Strominger:2013jfa} a remarkable connection was found between the asymptotic symmetries of gauge and gravity theories in flat space and the physics in the infrared. In particular, this has established \cite{Strominger:2013lka,Strominger:2013jfa,He:2014cra,Strominger:2014pwa,He} that Weinberg's soft theorems \cite{Weinberg:1965nx} for massless particles can be thought of as the Ward identity for infinite-dimensional global symmetries acting on the null-infinity in Minkowski space-time. Later this was extended to massive particles \cite{Kapec:2015ena,Campiglia:2015qka,Campiglia:2015kxa} in which case the infinite dimensional symmetries act on timelike infinity rather than null-infinity. In this paper we focus only on the massless case.

In four dimensions the space of null-momentum directions is a two-sphere - the so-called celestial sphere. The four dimensional Lorentz group $SL(2,\mathbb{C})$ acts on the celestial sphere as the group of global conformal transformations. Now in a related development \cite{Pasterski:2016qvg,Pasterski:2017kqt,Pasterski:2017ylz} Pasterski, Shao and Strominger have proposed an integral (Mellin) transform of massless (and massive) scattering amplitude (S-matrix) with the property that the resulting amplitude transforms under $SL(2,\mathbb{C})$ as the correlation function of products of $SL(2,\mathbb{C})$ primaries living on the celestial sphere. From the space-time point of view one gets the amplitude on the celestial sphere if, instead of the plane waves, one uses the conformal primary wave-functions \cite{Pasterski:2017kqt,Cheung:2016iub,deBoer:2003vf} to describe the external particles. The study of \cite{Pasterski:2017kqt} also established that the conformal primary wave-functions are closely related to the Unitary Principal Continuous Series Representation of $SL(2,\mathbb{C})$ (Lorentz group) \cite{bargman,Gadde:2017sjg,Hogervorst:2017sfd,Simmons-Duffin:2017nub}. The Mellin transformed amplitude has been computed for several field theories at both tree and loop level \cite{Pasterski:2017ylz,Cardona:2017keg,Lam:2017ofc,Banerjee:2017jeg,Schreiber:2017jsr}.

Now, one of the most attractive features of this approach \cite{Pasterski:2016qvg,Pasterski:2017kqt,Pasterski:2017ylz} is that it suggests a way of going \textit{directly} from the momentum space to null-infinity. The connection between soft theorems and asymptotic symmetries also suggests that one should be able to do this. But this particular approach goes beyond the soft sector of the gauge or gravity theory. In an earlier paper \cite{Banerjee:2018gce} we have studied this question from the Hilbert space perspective. Following \cite{Pasterski:2016qvg,Pasterski:2017kqt,Pasterski:2017ylz} we have constructed \cite{Banerjee:2018gce} a basis of massless single particle quantum states which transform in the unitary principal series representation of the Lorentz group. It turns out that these states also transform nicely under global space-time translations. From physical point of view a massless particle in this state can be thought of as living on the null-infinity. 

An interesting question to ask is that how one can see the connection between soft-theorems and symmetries in this approach. In this paper we take preliminary steps in this direction by rewriting Weinberg's soft theorems in this basis \cite{Pasterski:2016qvg,Pasterski:2017kqt,Pasterski:2017ylz,Banerjee:2018gce} and relating it to infinite dimensional global symmetries \cite{Strominger:2013lka,Strominger:2013jfa}. Technically this is very similar to \cite{Kapec:2017gsg} except that we are using a different basis. 

%Now for the sake of completeness let us briefly review part of \cite{Banerjee:2018gce} which is relevant for our purpose. But before that let us specify the conventions we use in this paper. 

\section{conventions}
In this paper we consider only massless particles. It is well known that we can associate a Hermitian matrix with a (null) four-vector $p$ in the following way, 

\be
p= 
\begin{pmatrix}
p^0 - p^3 & p^1 + i p^2 \\
p^1 - i p^2 & p^0 + p^3
\end{pmatrix} 
\ee

The determinant of the matrix gives the norm of the four vector, i.e, $\det p = -p^2$. Now let $\Lambda$ be a $SL(2,\mathbb{C})$ matrix which acts on $p$ as, 

\be\label{SL}
p \rightarrow p' = \Lambda p \Lambda^{\dagger} 
=
\begin{pmatrix}
p'^0 - p'^3 & p'^1 + i p'^2 \\
p'^1 - i p'^2 & p'^0 + p'^3
\end{pmatrix},  \ \
\Lambda =
\begin{pmatrix}
a & b\\
c & d
\end{pmatrix}\in SL(2,\mathbb{C})\ee

Under this action $p'$ is again an Hermitian matrix and the determinant does not change, i.e, $\det p = \det p'$ and so the the norm of the four-vectors $p$ and $p'$ are the same. So Eq-$\ref{SL}$ represents a Lorentz transformation. In this way we can associate to every $SL(2,\mathbb{C})$ matrix $\Lambda$ a Lorentz transformation which we also denote by $\Lambda$. To be more precise, the pair of $SL(2,\mathbb{C})$ matrices $\{\pm \Lambda\}$ get mapped to the same Lorentz transformation. In terms of the components of the four vector $p$ we can write the Lorentz transformation as, 

\be
p'^{\mu} = {\Lambda^{\mu}}_{\nu}(a,b,c,d) \ p^{\nu} 
\ee

where ${\Lambda^{\mu}}_{\nu}(a,b,c,d)$ is calculated from Eq-$\ref{SL}$. 

Another important quantity is the complex number $z$ defined as,

\be
z = \frac{p^1 + i p^2}{p^0 + p^3}, \ \ p^2 =0
\ee

Topologically a space-like cross-section of the null-cone is a two-sphere and $z$ is the stereographic coordinate on that. So $z$ specifies the direction of the null momentum vector. The space of null directions is also known as the celestial sphere. 

Under (Lorentz) $SL(2,\mathbb{C})$ transformation $z$ transforms as,

\be
z \rightarrow z' = \frac{p'^1 + i p'^2}{p'^0 + p'^3} = \frac{az+b}{cz+d}
\ee

So the Lorentz transformations act as global conformal transformations on $S^2$. 

In terms of $z$ a null vector $p$ can be parametrized as, 

\be
p = E \bigg(1, \frac{z+\bar z}{1 + z\bar z}, \frac{-i(z-\bar z)}{1 + z\bar z}, \frac{1-z\bar z}{1+z\bar z}\bigg), \ \  E = p^0 
\ee

An alternative parametrization is,

\be
p = \omega \bigg(1 + z\bar z, z+\bar z, -i(z-\bar z), 1-z\bar z\bigg)= \frac{E}{1+z\bar z} \bigg(1 + z\bar z, z+\bar z, -i(z-\bar z), 1-z\bar z\bigg)
\ee

We will use both. 

Under (Lorentz) $SL(2,\mathbb{C})$ transformation, 

\be
E \rightarrow E' = E \ \frac{|az+b|^2 + |cz+d|^2}{1 + z\bar z} 
\ee

In terms of $\omega$, 

\be
\omega \rightarrow \omega' = \omega |cz+d|^2 
\ee

Let us now say a few things about the unitary operators which implement Lorentz transformation in Hilbert space. Let us consider a Lorentz transformation $\Lambda$ which acts on space-time coordinates as, $x^{\mu}\rightarrow x'^{\mu} = {\Lambda^{\mu}}_{\nu} x^{\nu}$. Let $U(\Lambda)$ denote the corresponding unitary operator acting on the Hilbert space. For an infinitesimal transformation we write \cite{Weinberg:1995mt},

\be
U(\Lambda) = 1 + \frac{i}{2} \omega_{\mu\nu} J^{\mu\nu}+........, \ \ {\Lambda^{\mu}}_{\nu} = {\delta^{\mu}}_{\nu} + {\omega^{\mu}}_{\nu}, \ \  \omega_{\mu\nu} = \eta_{\mu\alpha} \ {\omega^{\alpha}}_{\nu} = - \omega_{\nu\mu}, \ \ J^{\mu\nu\dagger} = J^{\mu\nu}
\ee

Now the six generators $J^{\mu\nu}$ can be organised into rotation and boost three-vectors given by $\vec J= (J_1,J_2,J_3)$ and $\vec K = (K_1,K_2,K_3)$. $\vec J$ and $\vec K$ are both Hermitian. For our purpose it will be more convenient to write them in terms of the generators $(L_0,L_{\pm 1}, \bar L_0, \bar L_{\pm1})$ defined as \cite{Banerjee:2018gce},

\be
L_0 = \frac{-J_3 + i K_3 }{2} , \ \ \bar L_0 = - {L_0}^{\dagger}
\ee
\be
L_1 = \frac{-J_1 + i K_1}{2} + i \frac{-J_2 + i K_2}{2}, \ \ \bar L_1 = - {L_1}^{\dagger}
\ee
\be
L_{-1} = - \frac{- J_1 + i K_1}{2} + i \frac{- J_2 + i K_2}{2} , \ \  \bar L_{-1} = - {L_{-1}}^{\dagger}
\ee
The commutation relation between these generators can be written as, 
\be
[L_m , L_n] = (m-n) L_{m+n} , \ \  [\bar L_m , \bar L_n] = (m-n) \bar L_{m+n} , \ \ m, n = 0, \pm 1
\ee
The hermiticity of $\vec J$ and $\vec K$ gives the following reality condition,
\be
L_n^{\dagger} = - \bar L_n
\ee

Now consider an infinitesimal Lorentz transformation $U(\Lambda)$ generated by $L_n$ and $\bar L_n$. It can be written as

\be
U(\Lambda) = 1 + \alpha L_n + \bar\alpha\bar L_n + .......
\ee

where $\alpha$ is a complex number and $\bar\alpha$ is the complex conjugate of $\alpha$. Using the reality condition on $L_n$ and $\bar L_n$ it is easy to see that $U(\Lambda)$ is unitary. 

Now the $SL(2,\mathbb{C})$ transformations which correspond to $U(\Lambda)$ can be written in terms of $z$ as \cite{Banerjee:2018gce}

\be
z\rightarrow z' = z + \alpha z^n , \ \ n= 0,\pm1
\ee

So as usual $L_1$ generates translation, $L_0$ generates dilatation and $L_{-1}$ generates special conformal transformations on the two-sphere parametrized by $(z,\bar z)$.

\section{Review}
Let us now review the part of \cite{Pasterski:2016qvg,Pasterski:2017kqt,Pasterski:2017ylz,Banerjee:2018gce} which will be used in this paper.

Let us consider free massless particles in four dimensions. The Hilbert space of these particles carries unitary representation of the four dimensional Poincare group $ISO(3,1)$. For our purpose it will be useful to identify the Lorentz (sub-)group with $SL(2,\mathbb{C})$. 

Using this data one can construct an \textit{abstract three dimensional space-time} with coordinates denoted by $(u,z,\bar z)$ on which the Poincare group acts as,
\be\label{lorentz}
\Lambda (u,z, \bar z) =  \bigg( \frac{u \ (1+ z \bar z)}{|az+b|^2 + |cz+d|^2} \ , \frac{az+b}{cz+d} \ , \frac{\bar a \bar z + \bar b}{\bar c \bar z + \bar d}\bigg)
%\Lambda =
%\begin{pmatrix}
%a & b\\
%c & d
%\end{pmatrix}\in SL(2,\mathbb{C}), \ z\in\mathbb{C}
\ee 
\be
T(l) (u, z, \bar z) = (u + f(z, \bar z, l), z, \bar z)
\ee
where
\be\label{TR}
f(z, \bar z, l) = \frac{(l^0 - l^3) - (l^1 - i l^2) z - (l^1 + i l^2) \bar z + (l^0 + l^3) z \bar z}{1 + z \bar z}
\ee
Here $\Lambda$ is a $SL(2,\mathbb{C})$ (Lorentz) transformation and $T(l)$ is a global space-time translation by a four-vector $l^{\mu}$. There is no Poincare invariant metric in this space but rather there is a conformal structure.

This three dimensional space-time hosts free Heisenberg-picture creation and annihilation fields defined as,
\be
A^{\dagger}_{\lambda,\sigma} (u,z,\bar z) = \frac{1}{\sqrt{8\pi^4}} \bigg(\frac{1}{1+z \bar z}\bigg)^{1+i\lambda} \int_{0}^{\infty} dE \ E^{i\lambda} e^{iEu} \ a^{\dagger}(p,\sigma) , \ \lambda\in\mathbb{R},\ 2\sigma\in\mathbb{Z}
\ee 
\be
A_{\lambda,\sigma} (u,z,\bar z) = \frac{1}{\sqrt{8\pi^4}} \bigg(\frac{1}{1+z \bar z}\bigg)^{1-i\lambda} \int_{0}^{\infty} dE \ E^{-i\lambda} e^{-iEu} \ a(p,\sigma), \ \lambda\in\mathbb{R}, \ 2\sigma\in\mathbb{Z}
\ee
where $a^{\dagger}(p,\sigma)$ and $a(p,\sigma)$ are the momentum space creation and annihilation operators for a massless particle with null momentum $p$ and helicity $\sigma$. 

%In writing down the above equations we have assumed a specific parametrization of the null momentum given by,
%\be
%p = E \bigg(1, \frac{z+\bar z}{1 + z\bar z}, \frac{-i(z-\bar z)}{1 + z\bar z}, \frac{1-z\bar z}{1+z\bar z}\bigg) \rightarrow p^2 = 0
%\ee
The equations defining $A^{\dagger}_{\lambda,\sigma} (u,z,\bar z)$ and $A_{\lambda,\sigma} (u,z,\bar z)$ on the $(u,z,\bar z)$ space are analogous to the equations
\be
\phi^{-} (x^{\mu}) = \int\frac{d^3\vec p}{(2\pi)^3 2|\vec p|} e^{-ip.x} a^{\dagger}(p)
\ee
\be
\phi^{+} (x^{\mu}) = \int\frac{d^3\vec p}{(2\pi)^3 2|\vec p|} e^{ip.x} a(p)
\ee

which define, for example, the creation and annihilation parts of a free massless scalar (quantum) field on the Minkowski space-time. The only difference is that now we have defined $A^{\dagger}_{\lambda,\sigma} (u,z,\bar z)$ and $A_{\lambda,\sigma} (u,z,\bar z)$ also for spinning massless particles.

%The expression for $z$ in terms of the momentum components is given by, 
%\be
%z = \frac{p^1 + ip^2}{E + p^3} , \ \  -E^2 + (p^1)^2 + (p^2)^2 + (p^3)^2 =0
%\ee

%Now, as is well known, we can associate an Hermitian matrix with a (null) four-vector $p$ given by, 
%\be
%p= 
%\begin{pmatrix}
%p^0 - p^3 & p^1 + i p^2 \\
%p^1 - i p^2 & p^0 + p^3
%\end{pmatrix}
%\ee

Let us also define Lorentz tensors in this space as any function $\phi_{h,\bar h}(u,z,\bar z)$ which transforms under $SL(2,\mathbb{C})$ as, 
\be
\phi'_{h,\bar h} (u',z',\bar z') \bigg(\frac{dz'}{dz}\bigg)^{h} \bigg(\frac{d\bar z'}{d\bar z}\bigg)^{\bar h} = \phi_{h,\bar h} (u,z,\bar z)
\ee
where $(h,\bar h)$ are arbitrary complex numbers subject to the restriction that $h-\bar h \in \mathbb{Z}/2$. $(u',z',\bar z')$ are related to $(u,z,\bar z)$ by Eq-(\ref{lorentz}).

Under $SL(2,\mathbb{C})$ (Lorentz) transformation creation and annihilation fields transform homogeneously as tensors,
\be
U(\Lambda) \ A^{\dagger}_{\lambda,\sigma}(u, z,\bar z) \ U(\Lambda)^{-1} \\
 = \frac{1}{(cz+d)^{2h}} \frac{1}{(\bar c \bar z + \bar d)^{2 \bar h}}  \ A^{\dagger}_{\lambda,\sigma}\bigg(\frac{u \ (1+ z \bar z)}{|az+b|^2 + |cz+d|^2}, \frac{az+b}{cz+d} \ , \frac{\bar a \bar z + \bar b}{\bar c \bar z + \bar d}\bigg)
\ee
\be
U(\Lambda) \ A_{\lambda,\sigma}(u, z,\bar z) \ U(\Lambda)^{-1} \\
 = \frac{1}{(cz+d)^{2\bar h^{*}}} \frac{1}{(\bar c \bar z + \bar d)^{2 h^{*}}}  \ A_{\lambda,\sigma}\bigg(\frac{u \ (1+ z \bar z)}{|az+b|^2 + |cz+d|^2}, \frac{az+b}{cz+d} \ , \frac{\bar a \bar z + \bar b}{\bar c \bar z + \bar d}\bigg)
\ee
%Here $U(\Lambda)$ is the unitary representation of the Lorentz transformation $\Lambda$ acting on the Hilbert space and 
where $(h,\bar h)$ are little-group indices given by,
 \be
 h = \frac{1+ i\lambda - \sigma}{2} , \  \bar h = \frac{1+ i\lambda + \sigma}{2}
 \ee

$*$ denotes complex conjugation.  

Now under global space-time translation by a four-vector $l^{\mu}$ the fields transform as, 
\be
e^{-il.P} A^{\dagger}_{\lambda,\sigma}(u,z,\bar z) e^{il.P} = A^{\dagger}_{\lambda,\sigma} (u + f(z,\bar z,l), z, \bar z)
\ee
\be
e^{-il.P} A_{\lambda,\sigma}(u,z,\bar z) e^{il.P} = A_{\lambda,\sigma} (u + f(z,\bar z,l), z, \bar z)
\ee
 where $P^{\mu}$ is the translation operator and $f(z,\bar z,l)$ is given in Eq-$\ref{TR}$ .
 
 The (anti-)commutator of the creation and annihilation fields can be computed and is given by, 
  \be\label{CC}
 [A_{\lambda,\sigma} (u, z, \bar z), A^{\dagger}_{\lambda',\sigma'}(u', z', \bar z')]_{\pm}
= \frac{\delta_{\sigma \sigma'}}{2\pi} \frac{\Gamma\big(i(\lambda' - \lambda)\big)}{(1 + z \bar z)^{i(\lambda' - \lambda)}} \frac{\delta^2(z' -z)}{\big(-i(u'-u + i\epsilon)\big)^{i(\lambda' - \lambda)}}, \ \epsilon\rightarrow 0+
 \ee
 
Here we have used the relation,
\be
[a(p_1,\sigma_1),a^{\dagger}(p_2,\sigma_2)]_{\pm} = (2\pi)^3 2|\vec p_1| \delta^3({\vec p_1 - \vec p_2}) \delta_{\sigma_1 \sigma_2}
\ee
This commutator is Lorentz covariant and translation invariant.

This, at least in principle, complete the specification of the free field theory on $(u, z,\bar z)$ space because any other thing can be computed from the basic commutator given in Eq-($\ref{CC}$).

Now, as far as the action of the Poincare group is concerned, this abstract three dimensional space-time can be \textit{realized} as the null-infinity of the Minkowski space with $(u,z,\bar z)$ as Bondi coordinates. But this does not play any major role in the following discussions.

\section{Symmetries of free massless particles} 
The following discussion is strongly motivated by the remarkable connection between soft-theorems and asymptotic symmetries. In particular, it seems that if we want to use the approach of \cite{Pasterski:2016qvg,Pasterski:2017kqt,Pasterski:2017ylz,Banerjee:2018gce} to study the relation between soft-theorems and symmetries, then we cannot think of these symmetries as asymptotic. The simplest reason being that the starting point is not a theory defined in Minkowski space-time.

The following discussion is a very preliminary attempt in this direction. Loosely speaking we pick up the \textit{hard} part of the asymptotic symmetry generator (charge) and then turn off all interactions. In the non-interacting theory the hard-charge by itself is a symmetry generator and so we can study it in its own right in the free Hilbert space. This is also closer to the usual way of doing perturbative quantum field theory where we start with a free theory with certain symmetries and then ask what kind of interactions are allowed if the symmetry remains intact in the interacting theory. 

\subsection{(Super)translation}
Let us consider free massless particles of helicity $\sigma$. We keep the helicity unspecified because at this stage  nothing really depends crucially on helicity and the following discussion is purely algebraic. For simplicity let us just consider bosons. The conclusions remain unchanged for fermions.

The Hamiltonian can be written as, 
\be
H = \int d\mu(p) |\vec p| \ a^{\dagger}(p,\sigma) a(p,\sigma)
\ee
where $d\mu(p)$ is the Lorentz invariant measure given by
\be
d\mu(p) = \frac{d^3\vec p}{(2\pi)^3 2|\vec p|}
\ee
and 
\be
[a(p_1,\sigma),a^{\dagger}(p_2,\sigma)] = (2\pi)^3 \ 2|\vec p_1| \delta^3({\vec p_1 - \vec p_2}) 
\ee
%Let us now use the following parametrisation of the null momentum , 
%\be
%p = E \bigg(1, \frac{z+\bar z}{1 + z\bar z}, \frac{-i(z-\bar z)}{1 + z\bar z}, \frac{1-z\bar z}{1+z\bar z}\bigg) \rightarrow p^2 = 0
%\ee
Now parametrizing $p$ in terms of $(E,z,\bar z)$ we can write, 
\be
H = \int d\mu(p) \ E \ a^{\dagger}(E,z,\bar z,\sigma) a(E,z,\bar z,\sigma)
\ee
where
\be
d\mu(p) =  \frac{d^3\vec p}{(2\pi)^3 2|\vec p|} = \frac{E^2 dE}{(2\pi)^3 2E} \ \frac{4 d^2 z}{(1+ z\bar z)^2}, \ \ d^2 z = d\Re z \ d\Im z
\ee 

Let us now consider the following charges defined as,
\be
\boxed{
T_{f} = \int d\mu(p) \ E \ f(z,\bar z) a^{\dagger}(E,z,\bar z,\sigma) a(E,z,\bar z,\sigma) = T_f^{\dagger}}
\ee

where $f(z,\bar z)$ is an arbitrary real smooth function on the 2-sphere parametrized by $(z,\bar z)$. The operator $T_f$ has the following properties :

1)
\be
[H, T_f] = 0
\ee
So the charges are conserved for any function $f(z,\bar z)$. 

%2)
%\be
%T_f \ket{\Omega} = 0
%\ee
%where $\ket{\Omega}$ is the Fock vacuum.

2) 
\be
[T_f,T_{f'}] = 0 
\ee
for arbitrary $f$ and $f'$.

3)
The last important property is,
\be
[T_f, a^{\dagger}(E,z,\bar z,\sigma)] = E f(z,\bar z) a^{\dagger}(E,z,\bar z,\sigma) 
\ee
\be
[T_f, a(E,z,\bar z,\sigma)] = - E f(z,\bar z) a(E,z,\bar z,\sigma)
\ee

Let us now consider the unitary operator $U_f = e^{-i H_f}$. This unitary operator is a "Supertranslation - operator" in the $(u,z,\bar z)$ space. 

In order to see this let us go back to the basic creation/annihilation fields in the $(u,z,\bar z)$ space given by, 
\be
A^{\dagger}_{\lambda,\sigma} (u,z,\bar z) = \frac{1}{\sqrt{8\pi^4}} \bigg(\frac{1}{1+z \bar z}\bigg)^{1+i\lambda} \int_{0}^{\infty} dE \ E^{i\lambda} e^{iEu} \ a^{\dagger}(p,\sigma) , \ \lambda\in\mathbb{R},\ \sigma\in\mathbb{Z}
\ee 
\be
A_{\lambda,\sigma} (u,z,\bar z) = \frac{1}{\sqrt{8\pi^4}} \bigg(\frac{1}{1+z \bar z}\bigg)^{1-i\lambda} \int_{0}^{\infty} dE \ E^{-i\lambda} e^{-iEu} \ a(p,\sigma), \ \lambda\in\mathbb{R}, \ \sigma\in\mathbb{Z}
\ee

Now it is easy to see using the commutator of $T_f$ with the momentum space creation/annihilation operators that ,
\be
\begin{gathered}
e^{iT_f} A^{\dagger}_{\lambda,\sigma} (u,z,\bar z) e^{-iT_f} = \frac{1}{\sqrt{8\pi^4}} \bigg(\frac{1}{1+z \bar z}\bigg)^{1+i\lambda} \int_{0}^{\infty} dE \ E^{i\lambda} e^{iE(u + f(z,\bar z))} \ a^{\dagger}(p,\sigma) \\
= A^{\dagger}_{\lambda,\sigma} (u + f(z,\bar z),z,\bar z)
\end{gathered}
\ee
For infinitesimal transformation,
\be
f(z,\bar z)\frac{\partial A_{\lambda,\sigma}^{\dagger}}{\partial u} = i [T_f, A_{\lambda,\sigma}^{\dagger}]
\ee

Similarly, 
\be
\begin{gathered}
e^{iT_f} A_{\lambda,\sigma} (u,z,\bar z) e^{-iT_f} = \frac{1}{\sqrt{8\pi^4}} \bigg(\frac{1}{1+z \bar z}\bigg)^{1+i\lambda} \int_{0}^{\infty} dE \ E^{i\lambda} e^{iE(u + f(z,\bar z))} \ a(p,\sigma) \\
= A_{\lambda,\sigma} (u + f(z,\bar z),z,\bar z)
\end{gathered}
\ee
and 
\be
f(z,\bar z)\frac{\partial A_{\lambda,\sigma}}{\partial u} = i [T_f, A_{\lambda,\sigma}]
\ee 
for infinitesimal $f$.

So we can see that the unitary transformation $U_f = e^{-iT_f}$ generates a point transformation in the $(u,z,\bar z)$ space given by,
\be
(u,z,\bar z) \rightarrow (u+f(z,\bar z), z, \bar z)
\ee
where $f(z,\bar z)$ is an arbitrary real smooth function. Now if we choose $f(z,\bar z)$ to be of the following form
\be
f(z, \bar z, l) = \frac{(l^0 - l^3) - (l^1 - i l^2) z - (l^1 + i l^2) \bar z + (l^0 + l^3) z \bar z}{1 + z \bar z}
\ee
where $l^{\mu}$ is some four-vector, then we get back the global Minkowski space translations. If we relax this condition then we get "Supertranslations" in the $(u,z,\bar z)$ space. At this point we would like to emphasize that this has nothing to do with asymptotic symmetry. 

%Now since we know how the charges act on creation and annihilation operators in momentum space we can also also write down the action of the unitary operator $e^{-iT_f}$ on quantum fields defined on Minkowski space. Let us consider a massless scalar field in Minkowski space. The field corresponding to this is, 
%\be
%\phi(x) = \int d\mu(p) \big(e^{ip.x} a(p) + e^{-ip.x} a^{\dagger}(p) \big)
%\ee
%Now 
%\be
%\phi_f(x) = e^{iT_f} \phi(x) e^{-iT_f} = \int d\mu(p) \big(e^{ip.x} e^{i E_p f(z,\bar z)} a(p) + e^{-ip.x} e^{-iE_p f(z,\bar z)} a^{\dagger}(p) \big), \  \  E_p = |\vec p|
%\ee

Let us now see explicitly what happens to the commutator of $A_{\lambda,\sigma}$ and $A^{\dagger}_{\lambda,\sigma}$. The commutator is given by,
\be
 [A_{\lambda,\sigma} (u, z, \bar z), A^{\dagger}_{\lambda',\sigma'}(u', z', \bar z')]
= \frac{\delta_{\sigma \sigma'}}{2\pi} \frac{\Gamma\big(i(\lambda' - \lambda)\big)}{(1 + z \bar z)^{i(\lambda' - \lambda)}} \frac{\delta^2(z' -z)}{\big(-i(u'-u + i\epsilon)\big)^{i(\lambda' - \lambda)}}, \ \epsilon\rightarrow 0+
\ee

Now using the expressions for $e^{iT_f} A_{\lambda,\sigma} (u,z,\bar z) e^{-iT_f}$ and $e^{iT_f} A^{\dagger}_{\lambda,\sigma} (u,z,\bar z) e^{-iT_f}$ we get,

\be
\begin{gathered}
\[e^{iT_{f}} A_{\lambda,\sigma} (u, z, \bar z)e^{-iT_{f}}, e^{iT_{f}} A^{\dagger}_{\lambda',\sigma'}(u', z', \bar z')e^{-iT_{f}}\]  \\\\ 
= \[A_{\lambda,\sigma} (u + f(z,\bar z), z, \bar z), A^{\dagger}_{\lambda',\sigma'} (u+f(z',\bar z'), z', \bar z')\] \\\\
= \frac{\delta_{\sigma \sigma'}}{2\pi} \frac{\Gamma\big(i(\lambda' - \lambda)\big)}{(1 + z \bar z)^{i(\lambda' - \lambda)}} \frac{\delta^2(z' -z)}{\big(-i(u' + f(z',\bar z') -u - f(z,\bar z) + i\epsilon)\big)^{i(\lambda' - \lambda)}}  \\\\
= \frac{\delta_{\sigma \sigma'}}{2\pi} \frac{\Gamma\big(i(\lambda' - \lambda)\big)}{(1 + z \bar z)^{i(\lambda' - \lambda)}} \frac{\delta^2(z' -z)}{\big(-i(u' -u + i\epsilon)\big)^{i(\lambda' - \lambda)}} \\\\
= \[A_{\lambda,\sigma} (u, z, \bar z), A^{\dagger}_{\lambda',\sigma'} (u', z', \bar z')\]
\end{gathered}
\ee
The $f$'s in the denominator cancel due to the presence of $\delta^2(z-z')$. Physically this is just the fact that a free particle moves without changing its direction of motion and this makes the free theory on $(u,z,\bar z)$ space ultralocal in the spatial direction.

Let us now write down the commutation relation between the Lorentz generators and the supertranslations and show that the algebra is isomorphic to the BMS algebra without superrotations.

Let us start with the definition of the charge $T_f$ given by,

\be
T_{f} = \int d\mu(p) \ E \ f(z,\bar z) a^{\dagger}(E,z,\bar z,\sigma) a(E,z,\bar z,\sigma)
\ee

We will also need the following transformation properties,

\be
U(\Lambda)^{-1} a^{\dagger}(p,\sigma) U(\Lambda) = e^{i\sigma\theta(\Lambda,p)} a^{\dagger}(\Lambda^{-1}p,\sigma)
\ee
\be
U(\Lambda)^{-1} a(p,\sigma) U(\Lambda) = e^{-i\sigma\theta(\Lambda,p)} a(\Lambda^{-1}p,\sigma)
\ee

where the quantities appearing have their standard meaning \cite{Weinberg:1995mt}. Using this we can write, 

\be
\begin{gathered}
U(\Lambda)^{-1} T_f U(\Lambda) = \int d\mu(p) \ E \ f(z,\bar z) U(\Lambda)^{-1} a^{\dagger}(E,z,\bar z,\sigma) U(\Lambda)U(\Lambda)^{-1} a(E,z,\bar z,\sigma) U(\Lambda) \\
= \int d\mu(p) \ E \ f(z,\bar z) a^{\dagger}(\Lambda^{-1}E,\Lambda^{-1}z,\Lambda^{-1}\bar z,\sigma) a(\Lambda^{-1}E,\Lambda^{-1}z,\Lambda^{-1}\bar z,\sigma) \\
= \int d\mu(\Lambda^{-1}p) \ \Lambda\Lambda^{-1}E \ f(\Lambda\Lambda^{-1}z,\Lambda\Lambda^{-1}\bar z) a^{\dagger}(\Lambda^{-1}E,\Lambda^{-1}z,\Lambda^{-1}\bar z,\sigma) a(\Lambda^{-1}E,\Lambda^{-1}z,\Lambda^{-1}\bar z,\sigma) \\
= \int d\mu(p) \ \Lambda E \ f(\Lambda z,\Lambda\bar z) a^{\dagger}(E,z,\bar z,\sigma) a(E,z,\bar z,\sigma) \\
= \int d\mu(p) \ \bigg( E \ \frac{|az+b|^2 + |cz+d|^2}{1 + z\bar z}\bigg) \ f(\Lambda z,\Lambda\bar z) a^{\dagger}(E,z,\bar z,\sigma) a(E,z,\bar z,\sigma) \\
= \int d\mu(p) \ E \ \bigg( \frac{|az+b|^2 + |cz+d|^2}{1 + z\bar z} \ f(\Lambda z,\Lambda\bar z)\bigg) a^{\dagger}(E,z,\bar z,\sigma) a(E,z,\bar z,\sigma) \\
= \int d\mu(p) \ E \ f'(z,\bar z) a^{\dagger}(E,z,\bar z,\sigma) a(E,z,\bar z,\sigma) \\
= T_{f'}
\end{gathered}
\ee

So the supertranslation charges transform as, 

\be\label{BMS}
\boxed{
U(\Lambda)^{-1} T_f U(\Lambda) = T_{f'} , \ \ f'(z,\bar z) = \frac{|az+b|^2 + |cz+d|^2}{1 + z\bar z} \ f(\Lambda z,\Lambda\bar z)}
\ee

Here $\Lambda z = \frac{az+b}{cz+d}$.

Now since the global translation generators $P^{\mu}$ are already part of the supertranslations the above commutation relation already contains $U(\Lambda)^{-1} P^{\mu} U(\Lambda) = {\Lambda^{\mu}}_{\nu} P^{\nu}$. It is instructive to write the global translation generators in terms of $T_{f}$. In terms of creation/annihilation operators we can write, 

\be
P^{\mu} =  \int d\mu(p) \ p^{\mu} \ a^{\dagger}(E,z,\bar z,\sigma) a(E,z,\bar z,\sigma), 
\ee

Now,

\be
p = E \bigg(1, \frac{z+\bar z}{1 + z\bar z}, \frac{-i(z-\bar z)}{1 + z\bar z}, \frac{1-z\bar z}{1+z\bar z}\bigg)
\ee

So we get, 

\be
P^0 = H = T_f ,  \  \   f=1
\ee
\be
P^1 = T_f, \  \  f = \frac{z+\bar z}{1 + z\bar z}
\ee
\be
P^2 = T_f , \  \  f= \frac{-i(z-\bar z)}{1 + z\bar z}
\ee
\be
P^3 = T_f , \  \  f= \frac{1-z\bar z}{1+z\bar z}
\ee

Let us now write down the following linear combinations, 
\be
P^0 + P^3 = T_{00}, \  \  f = \frac{2z^0\bar z^0}{1 + z\bar z}
\ee
\be
P^0 - P^3 = T_{11},  \  \  f = \frac{2z^1\bar z^1}{1 + z\bar z}
\ee
\be
P^1 + i P^2 = T_{10}, \  \ f = \frac{2z^1\bar z^0}{1 + z\bar z}
\ee
\be
P^1 - i P^2 = T_{01}, \  \ f = \frac{2z^0\bar z^1}{1 + z\bar z}
\ee

\subsubsection{Global BMS}
Now to write down the generators of supertranslation we have to expand the function $f(z,\bar z)$ in a basis. The natural basis to do this is the spherical harmonics on the sphere \cite{Sachs:1962zza}. So we can define,

\be
T_{lm} = T_f , \  \  f \sim Y_{lm}(\theta,\phi)
\ee

We can use the basic relation Eq-$\ref{BMS}$ to calculate the commutator between the Lorentz generators and the $T_{lm}$. But for simplicity we will not do so. 

\subsubsection{Local BMS}
It is simpler to compute the commutator between Lorentz generators and $T_{pq}$ defined as \cite{Barnich:2009se}, 

\be
T_f = T_{pq} , \  \  f = \frac{2z^{p}\bar z^q}{1 + z\bar z}
\ee
where $(p,q)$ are integers. The corresponding generators will be denoted by $T_{pq}$. 

%One can see that for $p>1$ or $q>1$ the function $f$ has a singularity at infinity or south-pole of the sphere. This is why it is not globally defined. In our case, 

%\be
%T_{pq} \ket{E,z,\bar z,\sigma} = E \ \frac{2z^p \bar z^q}{1+z\bar z} \ket{E,z,\bar z,\sigma}
%\ee

%So except for a particle moving along the negative Z-axis the charge is finite. So let us find the commutator between $T_{pq}$ and the Lorentz generators. 

Let us also note that the reality condition on the generators is given by, 

\be
T_{pq}^{\dagger} = T_{qp}
\ee

So the Hermitian charge can be written as, 

\be
T = \sum_{p,q} C_{pq} T_{pq} , \  \  C_{pq}^{*}  = C_{qp}
\ee

Let us now write down the commutators between the Lorentz generators $(L_0,L_{\pm1},\bar L_0, \bar L_{\pm1})$ and $T_{pq}$. In order to this we have to use the relation derived earlier, 

\be
U(\Lambda)^{-1} T_f U(\Lambda) = T_{f'} , \ \ f'(z,\bar z) = \frac{|az+b|^2 + |cz+d|^2}{1 + z\bar z} \ f(\Lambda z,\Lambda\bar z)
\ee

Now it is much simpler to work with the function $F(z,\bar z)$ defined as, 

\be
F(z,\bar z) = (1+z \bar z) f(z,\bar z)
\ee

In terms of $F$, the charge can be written as,

\be
\begin{gathered}
T_{F} = T_f = \int d\mu(p) \ E \ f(z,\bar z) a^{\dagger}(E,z,\bar z,\sigma) a(E,z,\bar z,\sigma) \\
= \int d\mu(p) \ \frac{E}{1+z\bar z} \ \ (1+ z\bar z) f(z,\bar z) a^{\dagger}(E,z,\bar z,\sigma) a(E,z,\bar z,\sigma) \\
= \int d\mu(p) \ \frac{E}{1+z\bar z} \  F(z,\bar z) \ a^{\dagger}(E,z,\bar z,\sigma) a(E,z,\bar z,\sigma) 
\end{gathered}
\ee

Under (Lorentz) $SL(2,\mathbb{C})$ transformation $T_F (= T_f)$ transforms as, 

\be\label{F}
U(\Lambda)^{-1} T_F U(\Lambda) = T_{F'} , \ \ F'(z,\bar z) = |cz+d|^2 \ F(\Lambda z,\Lambda\bar z)
\ee

Now consider the infinitesimal $SL(2,\mathbb{C})$ transformation generated by $(L_n,\bar L_n)$, 

\be
U(\Lambda) = 1 + \alpha L_n + \bar\alpha \bar L_n , \  \  U(\Lambda)^{-1} = 1 - \alpha L_n - \bar\alpha \bar L_n
\ee

where $\bar\alpha$ is the complex conjugate of $\alpha$. In terms of the $z$ coordinate this generates the transformation,

\be
z \rightarrow \Lambda z = z + \alpha z^n
\ee

Now the generators $T_{pq}$ correspond to $F_{pq}(z,\bar z) = 2 z^p\bar z^q$, i.e, 

\be
\boxed{
T_{pq}=\int d\mu(p) \ \frac{E}{1+z\bar z} \  2z^p\bar z^q \ a^{\dagger}(E,z,\bar z,\sigma) a(E,z,\bar z,\sigma)= T_{qp}^{\dagger}}
\ee

Now we use the infinitesimal version of the relation given in Eq-$\ref{F}$ 

%with $F(z,\bar z)$ replaced by $F_{pq}(z,\bar z)$, 

\be
(1 - \alpha L_n - \bar\alpha \bar L_n) T_F (1 + \alpha L_n + \bar\alpha \bar L_n) = T_{F'}
\ee
\be
F'(z,\bar z) = \bigg(1-\frac{n+1}{2}(\alpha z^n + \bar\alpha\bar z^n) \bigg) F\big(z + \alpha z^n, \bar z + \bar\alpha\bar z^n\big)
\ee

and replace $F(z,\bar z)$ with $F_{pq}(z,\bar z) = 2z^p\bar z^q$ and $T_F$ with $T_{pq}$. Now after simple manipulations we get the following commutators, 

\be
\boxed{
[L_n , T_{pq}] = \bigg(\frac{n+1}{2} - p\bigg) T_{p+n,q} , \ \ [\bar L_n , T_{pq}] = \bigg(\frac{n+1}{2} - q\bigg) T_{p,q+n}, \ \ p,q\in{\mathbb{Z}}}
\ee

The rest of the commutators are, 
\be
\boxed{
[L_m,L_n] = (m-n) L_{m+n} , \  \  [\bar L_m, \bar L_n] = (m-n) \bar L_{m+n}, \  \ [T_{pq}, T_{p'q'}] = 0, \  \ [L_m, \bar L_n] = 0, \  \  L_n^{\dagger} = - \bar L_n}
\ee

Here $m,n = 0,\pm1$. 

This is a subalgebra of the extended BMS algebra, including superrotations, found in \cite{Barnich:2009se}.

\subsubsection{Space-time relaization}

Now since we know how the charges act on creation and annihilation operators in momentum space we can also also write down the action of the unitary operator $e^{-iT_f}$ on quantum fields defined on Minkowski space. Let us consider a massless scalar field in Minkowski space. The field corresponding to this is, 
\be
\phi(x) = \int d\mu(p) \big(e^{ip.x} a(p) + e^{-ip.x} a^{\dagger}(p) \big)
\ee
So the transformed field can be written as, 
\be
\phi_f(x) = e^{iT_f} \phi(x) e^{-iT_f} = \int d\mu(p) \big(e^{ip.x} e^{i E_p f(z,\bar z)} a(p) + e^{-ip.x} e^{-iE_p f(z,\bar z)} a^{\dagger}(p) \big), \  \  E_p = |\vec p|
\ee

Now if we choose $f(z,\bar z)$ to be of the following form, 
\be
f(z, \bar z, l) = \frac{(l^0 - l^3) - (l^1 - i l^2) z - (l^1 + i l^2) \bar z + (l^0 + l^3) z \bar z}{1 + z \bar z}
\ee
then 
\be
\phi_f(x^{\mu}) = \phi(x^{\mu} + l^{\mu})
\ee
So it is a space-time translation by four-vector $l^{\mu}$. But, there are infinitely many functions $f(z,\bar z)$ for which there is no such simple geometric interpretation in terms of space-time. In fact if we think of $a$ and $a^{\dagger}$ as complex conjugate c-numbers then $\phi_f$ is a solution of the massless Klein-Gordon equation if $\phi$ is. So it maps classical solutions to classical solutions as is expected for a symmetry transformation. 

Now suppose we demand that an interacting theory of massless scalar fields should also have this global symmetry. Arbitrary interaction will generically break this symmetry, but \cite{Strominger:2013jfa,He} have shown that if we couple the massless particle to dynamical gravity then this symmetry is preserved. This symmetry is now realized as the group of large diffeomorphisms and becomes the conventional BMS supertranslations. So now, as a part of a much bigger (gauge) symmetry group, the infinite dimensional global symmetry is realized locally geometrically in space-time. What this means is that we can write down actions, as integral over space-time of a local Lagrangian density, which are invariant under diffeomorphism and the infinite dimensional global symmetry is a subset of such diffeomorphisms. This should be contrasted with the situation in $(u,z,\bar z)$ space where supertranslations are realized locally from the very beginning. 

\subsection{Global U(1) Kac-Moody}

Now consider a charged massless scalar particle. The usual expression for the conserved charge of the particle can be written as, 
\be
Q = e \int d\mu(p) \big( a^{\dagger}(p) a(p) - b^{\dagger}(p)b(p) \big)
\ee
where the notation is standard.  The commutators are, 
\be
[Q, a^{\dagger}(p)] = e a^{\dagger}(p), \  \  [Q, a(p)] = -e a(p)
\ee
\be
[Q, b^{\dagger}(p)] = -e b^{\dagger}(p), \  \  [Q, b(p)] = e b(p)
\ee

Now just like supertranslation we can consider the following charges given by, 
\be
Q_f = e \int d\mu(p) f(z,\bar z) \big( a^{\dagger}(p) a(p) - b^{\dagger}(p)b(p) \big) = Q_f^{\dagger}
\ee
$Q_f$ is obviously conserved and $[Q_f,Q_{f'}]=0$ for all $f$. The commutator of the charge with the cretaion-annihilation operators is given by, 
\be
[Q_f, a^{\dagger}(p)] = e f(z,\bar z) a^{\dagger}(p), \  \  [Q_f, a(p)] = -e f(z,\bar z) a(p)
\ee
\be
[Q_f, b^{\dagger}(p)] = -e f(z,\bar z) a^{\dagger}(p), \  \  [Q_f, b(p)] = e f(z,\bar z) a(p)
\ee

On the creation and annihilation fields on the $(u,z,\bar z)$ space the unitary operator $e^{-iQ_f}$ acts like, 
\be
e^{-iQ_f} A_{\lambda}^{\dagger}(u,z,\bar z) e^{iQ_f} = e^{-ief(z,\bar z)} A_{\lambda}^{\dagger}(u,z,\bar z), \  \  e^{-iQ_f} A_{\lambda}(u,z,\bar z) e^{iQ_f} = e^{ief(z,\bar z)} A_{\lambda}(u,z,\bar z)
\ee
\be
e^{-iQ_f} B_{\lambda}^{\dagger}(u,z,\bar z) e^{iQ_f} = e^{ief(z,\bar z)} B_{\lambda}^{\dagger}(u,z,\bar z), \  \  e^{-iQ_f} B_{\lambda}(u,z,\bar z) e^{iQ_f} = e^{-ief(z,\bar z)} B_{\lambda}(u,z,\bar z)
\ee
So in the $(u,z,\bar z)$ charges act by local phase multiplication. So we get an infinite dimensional global $U(1)$ symmetry. There is one copy of $U(1)$ at every point of the sphere. 

%The phase depends on the spatial coordinates but not on the time-like coordinate. 

Let us now find out the Lorentz transformation properties of the charges. By the same method, that we have applied to supertranslation, we get,
\be\label{u1}
U(\Lambda)^{-1} Q_f U(\Lambda) = Q_{f'}, \  \  f'(z,\bar z) = f(\Lambda z, \Lambda\bar z)
\ee
So the total charge given by $Q_{f=1}$ is a scalar. Now we define the "moments",
\be
Q_{pq} = e \int d\mu(p) z^p\bar z^q \big( a^{\dagger}(p) a(p) - b^{\dagger}(p)b(p) \big) = Q_{qp}^{\dagger} , \  \  (p,q)\in\mathbb{Z}
\ee

Now as we have already discussed in the case of supertranslation, to get globally defined charges we should expand $f(z,\bar z)$ in speherical harmonics on the $S^2$. But it is easier to work with the above charges. This is also in the spirit of \cite{Strominger:2013lka}. 

The commutator of $Q_{pq}$ with the Lorentz generators $(L_0, L_{\pm1},\bar L_0,\bar L_{\pm1})$, as following from Eq-$\ref{u1}$ is given by, 
\be
\boxed{
[L_n, Q_{pq}] = - p \ Q_{p+n,q} \ , \  \  [\bar L_n, Q_{pq}] = - q \ Q_{p,q+n}  \ , \  \ [Q_{pq},Q_{p'q'}] =0} 
\ee

This is a level zero $U(1)$ Kac-Moody algebra except that we have confined ourselves to the global (Lorentz) conformal algebra. To connect it to the more familiar "holomorphic" and "antiholomorphic" charges, we can define $Q_p = Q_{p,q=0}$ and $\bar Q_{q} = Q_{p=0,q}$ which satisfy,
\be
[L_n, Q_p] = -p Q_{p+n} , \  \ [\bar L_n, Q_p] =0 , \ \ [Q_p,Q_q]=0
\ee
\be
[L_n, \bar Q_q] = 0 \ , \  \   [\bar L_n, \bar Q_q] = -q Q_{q+n}, \  \ [\bar Q_p, \bar Q_q] =0
\ee
\be
[Q, \bar Q] =0
\ee

One can repeat the same exercise with $f(z,\bar z)$ expanded in spherical harmonics. It will be interesting to do that.

The most important question of course is how constraining the symmetries are once we go to the interacting theory. For example the infinite dimensional global $U(1)$ symmetry of a free charged scalar particle is preserved if we couple it to photons \cite{He:2014cra,Strominger:2017zoo}. Then it is realized as large gauge transformations at null-infinity. So it is sufficient to have photons in the interacting theory. But is it also necessary ? 

\subsection{Discussion}
If we confine ourselves to the Hilbert space only then the statement that a free massless particle has infinite symmetries has, strictly speaking, no content. The Hilbert space of any free particle already has an infinite number of conserved quantities given by the number operators $a^{\dagger}(p)a(p)$. We can take any arbitrary linear combination of the number operators and generate another conserved quantity. A priori there is no reason to prefer one linear combination over the other. But we know that some of them are distinguished by the fact that they are related to space-time symmetries and are also symmetries of interacting theories. For example, Global space-time translations belong to this class but there are many more. So how do we decide which symmetries are useful ? 

The connection between soft-theorems and symmetries suggest that in the free Hilbert space of massless particles we consider hermitian operators of the form, 

\be
Q_K = \sum_{\sigma,\sigma'} \int\int d\mu(p) d\mu(p')  a(p,\sigma)^{\dagger} K(\sigma p,\sigma' p') a(p',\sigma')
\ee

where $K$ is a hermitian kernel which we assume to be local in momentum space, i.e, it is a sum of derivatives of a finite number of momentum space delta function. Now let us impose the following reasonable conditions : 

1) The hermitian operators $Q_K$ corresponding to different $K$'s should form a closed algebra under commutation. The algebra may be finite or infinite dimensional. In the two examples we have studied they are infinite dimensional.

2) It seems that as far as the free theory is concerned we can think in terms of (quantum) fields either on Minkowski space or on the $(u,z,\bar z)$ space. The action of $Q_K$ on the fields should be local in \textit{at least} one description. Supertranslations are examples of this. The action of $T_f$ for a generic $f$ is local only in the $(u,z,\bar z)$ space whereas global translations are local in both descriptions. 

$Q_K$'s satisfying at least the above two conditions may reasonably be said to form a symmetry algebra of a free massless particle. Now, heuristically speaking, free particles are maximally symmetric objects. So classifying all possible kernels $K$ subject to the constraint that the charges $Q_K$ satisfy the above two conditions can possibly give us the global symmetries that a putative interacting theory of massless particles can have. This is a mathematically well-posed problem and it will be interesting to know the answer.

\section{S-Matrix and Soft Limits}

\subsection{A Convenient Set Of Variables}
In practice it is much simpler to write the creation and annihilation fields in terms of the new variables defined as, 

\be
U = u(1+z\bar z) , \  \omega = \frac{E}{1+z\bar z}
\ee

Under $SL(2,\mathbb{C})$ (Lorentz) transformation

\be
\Lambda (U,z,\bar z) = \bigg( \frac{U}{|cz+d|^2}, \ \frac{az+b}{cz+d}, \ \frac{\bar a \bar z + \bar b}{\bar c \bar z + \bar d} \bigg)
\ee
and under space-time translation by a 4-vector $l^{\mu}$,
\be
T(l) (u,z,\bar z) = (u + g(z,\bar z,l),z,\bar z)
\ee
where 
\be
g(z,\bar z,l) = (l^0 - l^3) - (l^1 - i l^2) z - (l^1 + i l^2) \bar z + (l^0 + l^3) z \bar z = (1 + z\bar z) f(z,\bar z,l)
\ee
In terms of these new variables we can write, 
\be
A^{\dagger}_{\lambda,\sigma} (U,z,\bar z) = \frac{1}{\sqrt{8\pi^4}} \int_{0}^{\infty} d\omega \ \omega^{i\lambda} e^{i\omega U} \ a^{\dagger}(p,\sigma)
\ee
and
\be
A_{\lambda,\sigma} (U,z,\bar z) = \frac{1}{\sqrt{8\pi^4}}\int_{0}^{\infty} d\omega \ \omega^{-i\lambda} e^{-i\omega U} \ a(p,\sigma)
\ee
The null momentum $p$ is now parametrized as,
\be
p = \omega (1+ z\bar z, z + \bar z, -i(z-\bar z), 1- z\bar z) = \frac{E}{1 + z\bar z} (1+ z\bar z, z + \bar z, -i(z-\bar z), 1- z\bar z)
\ee

%\be
%a(p,\sigma) = \sqrt{2\pi^2} \ \omega^{i\lambda} \int_{-\infty}^{\infty} dU \ e^{i\omega U} A_{\lambda,\sigma}(U,z,\bar z)
%\ee 

%and

%\be
%a^{\dagger}(p,\sigma) = \sqrt{2\pi^2} \ \omega^{-i\lambda} \int_{-\infty}^{\infty} dU \ e^{-i\omega U} A^{\dagger}_{\lambda,\sigma}(U,z,\bar z)
%\ee 

%Let us now express the S-matrix in terms of the basis states $\ket{\lambda,\sigma,u,z,\bar z}$. We write 

%\be
%S_{N\leftarrow M}\big(\{p_i,\sigma_i,n_i\}, out \ \big| \ \{k_j,\sigma_j,n_j\}, in \big) = \bra{\Omega} \prod_{i=M+1}^{M+N} a_{out}(p_i,\sigma_i,n_i)\prod_{j=1}^{M} a^{\dagger}_{in}(k_j,\sigma_j,n_j) \ket{\Omega}
%\ee

%\be
%a_{in/out}(p,\sigma,n) = \sqrt{2\pi^2} \ \omega^{i\lambda} \int_{-\infty}^{\infty} dU \ e^{i\omega U} A_{\lambda,\sigma,n}^{in/out}(U,z,\bar z)
%\ee

%\be
%a_{in/out}^{\dagger}(p,\sigma,n) = \sqrt{2\pi^2} \ \omega^{-i\lambda} \int_{-\infty}^{\infty} dU \ e^{-i\omega U} A^{\dagger in/out}_{\lambda,\sigma,n}(U,z,\bar z)
%\ee 

%\be
%\begin{aligned}
%S_{N\leftarrow M}\big(\{p_i,\sigma_i,n_i\}, out \ \big| \ \{k_j,\sigma_j,n_j\}, in \big) \\
%= (\sqrt{2\pi^2})^{M+N} \bigg(\prod_{i=M+1}^{M+N}\omega_i^{i\lambda_i}\int_{-\infty}^{\infty}dU_i e^{i\omega_i U_i}\bigg)\bigg(\prod_{j=1}^{M}\omega_j^{-i\lambda_j}\int_{-\infty}^{\infty}dU_j e^{-i\omega_j U_j}\bigg) \\
 %\bra{\Omega} \prod_{i=M+1}^{M+N} A^{out}_{n_i,\lambda_i,\sigma_i}(U_i,z_i,\bar z_i) \prod_{j=1}^{M}A^{\dagger in}_{n_j,\lambda_j,\sigma_j}(U_j,z_j,\bar z_j) \ket{\Omega}
%\end{aligned}
%\ee \\

\subsection{S-matrix}
Let us now consider a $M\rightarrow N$ massless particle scattering. The quantum state of a particle is specified by its null four-momentum $p$, helicity $\sigma$ and an extra label $n$ which carries information about other internal quantum numbers like charge. The corresponding creation and annihilation operators are denoted by $a^{\dagger}(p,\sigma,n)$ and $a(p,\sigma,n)$. Since we are considering a scattering process, there are also the labels $in$ and $out$ depending on whether the particle is incoming or outgoing.

Let us now introduce the $in$ and $out$ creation and annihilation free-fields in the $(U,z,\bar z)$ space, defined as   
\be
A^{\dagger in/out}_{n,\lambda,\sigma} (U,z,\bar z) = \frac{1}{\sqrt{8\pi^4}} \int_{0}^{\infty} d\omega \ \omega^{i\lambda} e^{i\omega U} \ a^{\dagger}_{in/out}(p,\sigma,n)
\ee
\be
A^{in/out}_{n,\lambda,\sigma} (U,z,\bar z) = \frac{1}{\sqrt{8\pi^4}}\int_{0}^{\infty} d\omega \ \omega^{-i\lambda} e^{-i\omega U} \ a_{in/out}(p,\sigma,n)
\ee

Using these we can write,
\be\label{SU}
\begin{gathered}
\tilde{A}\big(\{n,\lambda,\sigma, U,z,\bar z\}\big) =
\bra{\Omega} \prod_{i=M+1}^{M+N} A^{out}_{n_i,\lambda_i,\sigma_i}(U_i,z_i,\bar z_i) \prod_{j=1}^{M}A^{\dagger in}_{n_j,\lambda_j,\sigma_j}(U_j,z_j,\bar z_j) \ket{\Omega} \\
=\frac{1}{(\sqrt{8\pi^4})^{M+N}} \bigg(\prod_{i=M+1}^{M+N}\int_{0}^{\infty}d\omega_i \omega_i^{- i\lambda_i}e^{-i\omega_i U_i}\bigg)\bigg(\prod_{j=1}^{M}\int_{0}^{\infty}d\omega_j \omega_j^{i\lambda_j} e^{i\omega_j U_j}\bigg) \\
S_{N\leftarrow M}\big(\{p_i,\sigma_i,n_i\}, out \ \big| \ \{p_j,\sigma_j,n_j\}, in \big)
\end{gathered}
\ee
where $\ket{\Omega}$ is the Poincare vacuum and the S-matrix is defined as usual by, 
\be
S_{N\leftarrow M}\big(\{p_i,\sigma_i,n_i\}, out \ \big| \ \{p_j,\sigma_j,n_j\}, in \big) = \bra{\Omega}\prod_{i=M+1}^{M+N} a_{out}(p_i,\sigma_i,n_i) \prod_{j=1}^{M} a^{\dagger}_{in}(p_j,\sigma_j,n_j)\ket{\Omega}
\ee

We will refer to L.H.S of Eq-$\ref{SU}$ as the $\tilde A$-amplitude. This is essentially the integral transform of the S-matrix and its generalization which appears in \cite{Pasterski:2016qvg,Pasterski:2017kqt,Pasterski:2017ylz,Banerjee:2018gce}. As we will see this form of the L.H.S is more convenient for discussing the soft limit. 

It will be interesting to construct the interpolating fields on this space but this is beyond the scope of the present paper.

\subsection{Soft Limit}
Let us now consider a $M\rightarrow N+1$ scattering process. We have added an extra particle in the final state with momentum $p$ and helicity $\sigma$ which will be taken to be soft at the end. The corresponding $\tilde A$-amplitude can be written as,
\be
\begin{gathered}
\bra{\Omega} A^{out}_{\lambda,\sigma}(U,z,\bar z) \prod_{i=M+1}^{M+N} A^{out}_{n_i,\lambda_i,\sigma_i}(U_i,z_i,\bar z_i) \prod_{j=1}^{M}A^{\dagger in}_{n_j,\lambda_j,\sigma_j}(U_j,z_j,\bar z_j) \ket{\Omega} \\
=\frac{1}{(\sqrt{8\pi^4})^{M+N}} \bigg(\frac{1}{\sqrt{8\pi^4}}\int_{0}^{\infty} d\omega \omega^{-i\lambda} e^{-i\omega U} \prod_{i=M+1}^{M+N}\int_{0}^{\infty}d\omega_i \omega_i^{- i\lambda_i}e^{-i\omega_i U_i}\bigg)\bigg(\prod_{j=1}^{M}\int_{0}^{\infty}d\omega_j \omega_j^{i\lambda_j} e^{i\omega_j U_j}\bigg) \\
S_{N+1\leftarrow M}\big(\{p,\sigma \ ; \ p_i,\sigma_i,n_i\}, out \ \big| \ \{p_j,\sigma_j,n_j\}, in \big)
\end{gathered}
\ee
In the definition of the $\tilde A$-amplitude all the external legs of the S-matrix are integrated over. So in order to take the soft limit we have to make the leg corresponding to the would-be soft particle free. This can be done by inverting the transformation at the soft leg. The resulting formula can be written as,
\be
\begin{gathered}
\sqrt{2\pi^2} \int_{-\infty}^{\infty}d\lambda \ \omega^{i\lambda -1} e^{i\omega U} \bra{\Omega} A^{out}_{\lambda,\sigma}(U,z,\bar z) \prod_{i=M+1}^{M+N} A^{out}_{n_i,\lambda_i,\sigma_i}(U_i,z_i,\bar z_i) \prod_{j=1}^{M}A^{\dagger in}_{n_j,\lambda_j,\sigma_j}(U_j,z_j,\bar z_j) \ket{\Omega} \\
=\frac{1}{(\sqrt{8\pi^4})^{M+N}} \bigg(\prod_{i=M+1}^{M+N}\int_{0}^{\infty}d\omega_i \omega_i^{- i\lambda_i}e^{-i\omega_i U_i}\bigg)\bigg(\prod_{j=1}^{M}\int_{0}^{\infty}d\omega_j \omega_j^{i\lambda_j} e^{i\omega_j U_j}\bigg) \\
S_{N+1\leftarrow M}\big(\{p,\sigma \ ; \ p_i,\sigma_i,n_i\}, out \ \big| \ \{p_j,\sigma_j,n_j\}, in \big)
\end{gathered}
\ee

We can see that the outgoing leg of the S-matrix with momentum $p$ and helicity $\sigma$ is now free. Now one can take the soft limit by multiplying both sides with appropriate factors of $\omega$ and taking $\omega$ to zero. Let us denote this factor by $f(\omega)$.

\be\label{sl}
\begin{gathered}
\lim_{\omega\rightarrow 0+} f(\omega)\sqrt{2\pi^2} \int_{-\infty}^{\infty}d\lambda \ \omega^{i\lambda -1} e^{i\omega U} \bra{\Omega} A^{out}_{\lambda,\sigma}(U,z,\bar z) \prod_{i=M+1}^{M+N} A^{out}_{n_i,\lambda_i,\sigma_i}(U_i,z_i,\bar z_i) \prod_{j=1}^{M}A^{\dagger in}_{n_j,\lambda_j,\sigma_j}(U_j,z_j,\bar z_j) \ket{\Omega} \\
=\frac{1}{(\sqrt{8\pi^4})^{M+N}} \bigg(\prod_{i=M+1}^{M+N}\int_{0}^{\infty}d\omega_i \omega_i^{- i\lambda_i}e^{-i\omega_i U_i}\bigg)\bigg(\prod_{j=1}^{M}\int_{0}^{\infty}d\omega_j \omega_j^{i\lambda_j} e^{i\omega_j U_j}\bigg) \\
\lim_{\omega\rightarrow 0+} f(\omega) S_{N+1\leftarrow M}\big(\{p,\sigma \ ; \ p_i,\sigma_i,n_i\}, out \ \big| \ \{p_j,\sigma_j,n_j\}, in \big)
\end{gathered}
\ee

We will this formula to study the soft-limit.

\section{Leading Soft Photon and Graviton}

For leading soft photon and graviton we have to take $f(\omega) = \omega$. With this the Eq-$\ref{sl}$ becomes, 
\be\label{SP}
\begin{gathered}
\lim_{\omega\rightarrow 0+}\sqrt{2\pi^2} \int_{-\infty}^{\infty}d\lambda \ \omega^{i\lambda} e^{i\omega U} \bra{\Omega} A^{out}_{\lambda,\sigma}(U,z,\bar z) \prod_{i=M+1}^{M+N} A^{out}_{n_i,\lambda_i,\sigma_i}(U_i,z_i,\bar z_i) \prod_{j=1}^{M}A^{\dagger in}_{n_j,\lambda_j,\sigma_j}(U_j,z_j,\bar z_j) \ket{\Omega} \\
=\frac{1}{(\sqrt{8\pi^4})^{M+N}} \bigg(\prod_{i=M+1}^{M+N}\int_{0}^{\infty}d\omega_i \omega_i^{- i\lambda_i}e^{-i\omega_i U_i}\bigg)\bigg(\prod_{j=1}^{M}\int_{0}^{\infty}d\omega_j \omega_j^{i\lambda_j} e^{i\omega_j U_j}\bigg) \\
\lim_{\omega\rightarrow 0+} \omega S_{N+1\leftarrow M}\big(\{p,\sigma \ ; \ p_i,\sigma_i,n_i\}, out \ \big| \ \{p_j,\sigma_j,n_j\}, in \big)
\end{gathered}
\ee

Now consider the operator 
\be
J_{\sigma, \omega}(z,\bar z) = \sqrt{2\pi^2} \ \int_{-\infty}^{\infty}d\lambda \ \omega^{i\lambda} e^{i\omega U}A^{out}_{\lambda,\sigma}(U,z,\bar z) = \omega a^{out}(p,\sigma)
\ee
We have omitted the $U$ dependence in $J_{\sigma,\omega}$ because it is equal to the $U$-independent annihilation operator $\omega a^{out}(p,\sigma)$. We can also work by keeping the $U$ dependence and then the soft theorems will imply that the $U$ dependence is trivial. In fact this is more appropriate from our point of view. We discuss this in detail in a later section.

Now using the $SL(2,\mathbb{C})$ transformation property of $A_{\lambda,\sigma}$ we get,
\be
U(\Lambda) J_{\sigma,\omega}(z,\bar z) U(\Lambda)^{-1} = \frac{1}{(cz+d)^{2h}}\frac{1}{(\bar c \bar z + \bar d)^{2\bar h}} J_{\sigma, \omega |cz+d|^2}\bigg(\frac{az+b}{cz+d}, \ \frac{\bar a \bar z + \bar b}{\bar c \bar z + \bar d}\bigg)
\ee
where
\be
h = \frac{1+\sigma}{2}, \ \bar h = \frac{1-\sigma}{2}
\ee
Now if we take the soft limit $\omega\rightarrow 0+$ we can define the soft-operator \cite{Kapec:2017gsg} $J_{\sigma}(z,\bar z)= J_{\sigma, \omega=0}(z,\bar z)$ whose Lorentz transformation property is
\be
U(\Lambda) J_{\sigma}(z,\bar z) U(\Lambda)^{-1} = \frac{1}{(cz+d)^{1+\sigma}}\frac{1}{(\bar c \bar z + \bar d)^{1-\sigma}} J_{\sigma}\bigg(\frac{az+b}{cz+d}, \ \frac{\bar a \bar z + \bar b}{\bar c \bar z + \bar d}\bigg)
\ee

We will now write Weinberg's soft theorems in terms of the soft operator $J_{\sigma}$ for $\sigma = +1$ and $\sigma=+2$.

%We will discuss the "soft operator" from a slightly different point of view in a later section.

%We can also study the transformation of $J_{\sigma,\omega}(U,z,\bar z)$ under global translation by a four vector $l^{\mu}$. The transformation property is given by,
%\be
%e^{-il.P} J_{\sigma,\omega} (U,z,\bar z) e^{il.P} = e^{-i\omega g(z,\bar z, l)} J_{\sigma,\omega} (U + g(z,\bar z,l),z,\bar z)
%\ee
%where
%\be
%g(z,\bar z,l) = (l^0 - l^3) - (l^1 - i l^2) z - (l^1 + i l^2) \bar z + (l^0 + l^3) z \bar z
%\ee
%So in the soft limit $\omega\rightarrow 0+$ we get,
%\be
%e^{-il.P} J_{\sigma} (U,z,\bar z) e^{il.P} = J_{\sigma} (U + g(z,\bar z,l),z,\bar z)
%\ee

%We will see that soft theorems imply that the $U$ dependence of the soft-operator is trivial.

%We can see that $\lambda$ dependence drops out from the $SL(2,\mathbb{C})$ (Lorentz) transformation property. This is the reason that we have not introduced any $\lambda$ dependence in $J_{\sigma}(\omega,z,\bar z)$.

\subsection{Soft Photon Limit}
For photons $\sigma=\pm1$ and we denote the corresponding soft photon operators by $J_{\pm}(z,\bar z)$ whose $SL(2,\mathbb{C})$ transformation property is given by,

\be
U(\Lambda) J_{+}(z,\bar z) U(\Lambda)^{-1} = \frac{1}{(cz+d)^{2}} \ J_{+} \bigg(\frac{az+b}{cz+d}, \frac{\bar a \bar z + \bar b}{\bar c \bar z + \bar d}\bigg), \  \sigma=+1
\ee
and 
\be
U(\Lambda) J_{-}(z,\bar z) U(\Lambda)^{-1} = \frac{1}{(\bar c \bar z+\bar d)^{2}} \ J_{-} \bigg(\frac{az+b}{cz+d}, \frac{\bar a \bar z + \bar b}{\bar c \bar z + \bar d}\bigg), \ \sigma=-1
\ee

So $J_{\pm}$ transform like $(1,0)$ and $(0,1)$ (quasi-) primary operators of a 2-D CFT. 

Let us now write down Weinberg's soft-photon theorem in terms of $J_{\pm}$.

Let us consider a positive helicity ($\sigma=+1$) outgoing soft photon. In this case Weinberg's soft-photon theorem reduces to \cite{He:2014cra,Strominger:2017zoo},
\be
\begin{gathered}
\lim_{\omega\rightarrow 0+} \omega S_{N+1\leftarrow M}\big(\{p,\sigma = +1 \ ; \ p_i,\sigma_i,n_i\}, out \ \big| \ \{p_j,\sigma_j,n_j\}, in \big) \\
= \frac{1}{\sqrt 2} \bigg[ \sum_{i=M+1}^{M+N}\frac{Q_i}{z-z_i} - \sum_{j=1}^{M} \frac{Q_j}{z-z_j}\bigg] S_{N\leftarrow M}\big(\{\ p_i,\sigma_i,n_i\}, out \ \big| \ \{p_j,\sigma_j,n_j\}, in \big)
\end{gathered}
\ee

where $Q_p$ is the charge of the $p$-th particle. Substituting this in Eq-$\ref{SP}$ we get the following relation,
\be
\begin{gathered}
\lim_{\omega\rightarrow 0+}\sqrt{2\pi^2} \ \int_{-\infty}^{\infty}d\lambda \ \omega^{i\lambda} e^{i\omega U} \bra{\Omega} A^{out}_{\lambda,\sigma=+1}(U,z,\bar z) \prod_{i=M+1}^{M+N} A^{out}_{n_i,\lambda_i,\sigma_i}(U_i,z_i,\bar z_i) \prod_{j=1}^{M}A^{\dagger in}_{n_j,\lambda_j,\sigma_j}(U_j,z_j,\bar z_j) \ket{\Omega} \\
= \frac{1}{\sqrt 2} \bigg[ \sum_{i=M+1}^{M+N}\frac{Q_i}{z-z_i} - \sum_{j=1}^{M} \frac{Q_j}{z-z_j}\bigg] \bra{\Omega}\prod_{i=M+1}^{M+N} A^{out}_{n_i,\lambda_i,\sigma_i}(U_i,z_i,\bar z_i) \prod_{j=1}^{M}A^{\dagger in}_{n_j,\lambda_j,\sigma_j}(U_j,z_j,\bar z_j) \ket{\Omega}
\end{gathered}
\ee

In terms of the soft-photon operator $J_{+}(z,\bar z)$ Weinberg's soft-photon theorem takes the following form, 

\be
\boxed{
\begin{gathered}
\bra{\Omega} J_{+}(z,\bar z) \prod_{i=M+1}^{M+N} A^{out}_{n_i,\lambda_i,\sigma_i}(U_i,z_i,\bar z_i) \prod_{j=1}^{M}A^{\dagger in}_{n_j,\lambda_j,\sigma_j}(U_j,z_j,\bar z_j) \ket{\Omega} \\
= \frac{1}{\sqrt 2} \bigg[ \sum_{i=M+1}^{M+N}\frac{Q_i}{z-z_i} - \sum_{j=1}^{M} \frac{Q_j}{z-z_j}\bigg] \bra{\Omega}\prod_{i=M+1}^{M+N} A^{out}_{n_i,\lambda_i,\sigma_i}(U_i,z_i,\bar z_i) \prod_{j=1}^{M}A^{\dagger in}_{n_j,\lambda_j,\sigma_j}(U_j,z_j,\bar z_j) \ket{\Omega}
\end{gathered}}
\ee

%Now we can see that the R.H.S of the above equation does not depend on $U$ and we can write $\partial_{U} J_{+}(U,z,\bar z) =0$ as an operator equation because its matrix elements between arbitrary in and out states vanish. So without any loss of generality we can simply write the soft-photon operator as $J_{+}(z,\bar z)$. In terms of this we get,

%\be
%\boxed{
%\begin{gathered}
%\bra{\Omega} J_{+}(z,\bar z) \prod_{i=M+1}^{M+N} A^{out}_{n_i,\lambda_i,\sigma_i}(U_i,z_i,\bar z_i) \prod_{j=1}^{M}A^{\dagger in}_{n_j,\lambda_j,\sigma_j}(U_j,z_j,\bar z_j) \ket{\Omega} \\
%= \frac{1}{\sqrt 2} \bigg[ \sum_{i=M+1}^{M+N}\frac{Q_i}{z-z_i} - \sum_{j=1}^{M} \frac{Q_j}{z-z_j}\bigg] \bra{\Omega}\prod_{i=M+1}^{M+N} A^{out}_{n_i,\lambda_i,\sigma_i}(U_i,z_i,\bar z_i) \prod_{j=1}^{M}A^{\dagger in}_{n_j,\lambda_j,\sigma_j}(U_j,z_j,\bar z_j) \ket{\Omega}
%\end{gathered}}
%\ee

The leading soft photon theorem appears in this form in \cite{Strominger:2013lka} written in a different basis of states. This has the form of a $U(1)$ Kac-Moody ward identity in a 2-D CFT \cite{Strominger:2013lka,He:2014cra,Cheung:2016iub} except that now the fields depend on an extra coordinate given by $U$. But the $U$ direction does not play any role in this case. 

%This also shows that 
%\be
%\bar\partial J_{+} = 0
%\ee

\subsection{Soft Graviton Limit}

Let us now consider a positive helicity ($\sigma=+2$) outgoing soft-graviton. In this case Weinberg's soft graviton theorem can be cast in the following form \cite{He},
\be
\begin{gathered}
\lim_{\omega\rightarrow 0+} \omega S_{N+1\leftarrow M}\big(\{p,\sigma = +2 \ ; \ p_i,\sigma_i,n_i\}, out \ \big| \ \{p_j,\sigma_j,n_j\}, in \big) \\
= \frac{\kappa}{2}\bigg[ \sum_{j=1}^{M}\frac{\omega_j(\bar z- \bar z_j)}{z-z_j} - \sum_{i=M+1}^{M+N}\frac{\omega_i(\bar z- \bar z_i)}{z-z_i}\bigg] S_{N\leftarrow M}\big(\{\ p_i,\sigma_i,n_i\}, out \ \big| \ \{p_j,\sigma_j,n_j\}, in \big)
\end{gathered}
\ee

where $\kappa = \sqrt{32\pi G_N}$. We have also parametrized an arbitrary null momentum $p_{\alpha}$ as, 
\be
p_{\alpha} = \omega_{\alpha}\big(1+z_{\alpha}\bar z_{\alpha} , z_{\alpha}+\bar z_{\alpha}, -i(z_{\alpha}-\bar z_{\alpha}), 1-z_{\alpha}\bar z_{\alpha}\big)
\ee

Now substituting this in Eq-$\ref{SP}$ we get

\be\label{SG}
\begin{gathered}
\lim_{\omega\rightarrow 0+} \sqrt{2\pi^2} \int_{-\infty}^{\infty}d\lambda \ \omega^{i\lambda} e^{i\omega U} \bra{\Omega} A^{out}_{\lambda,\sigma = +2}(U,z,\bar z) \prod_{i=M+1}^{M+N} A^{out}_{n_i,\lambda_i,\sigma_i}(U_i,z_i,\bar z_i) \prod_{j=1}^{M}A^{\dagger in}_{n_j,\lambda_j,\sigma_j}(U_j,z_j,\bar z_j) \ket{\Omega} \\
=\frac{1}{(\sqrt{8\pi^4})^{M+N}} \bigg(\prod_{i=M+1}^{M+N}\int_{0}^{\infty}d\omega_i \omega_i^{- i\lambda_i}e^{-i\omega_i U_i}\bigg)\bigg(\prod_{j=1}^{M}\int_{0}^{\infty}d\omega_j \omega_j^{i\lambda_j} e^{i\omega_j U_j}\bigg) \\
\frac{\kappa}{2}\bigg[ \sum_{j=1}^{M}\frac{\omega_j(\bar z- \bar z_j)}{z-z_j} - \sum_{i=M+1}^{M+N}\frac{\omega_i(\bar z- \bar z_i)}{z-z_i}\bigg] S_{N\leftarrow M}\big(\ p_i,\sigma_i,n_i\}, out \ \big| \ \{p_j,\sigma_j,n_j\}, in \big)
\end{gathered}
\ee

Now in the same way we define the positive helicity soft-graviton operator $O_{+}(U,z,\bar z)$ by
\be
O_{+} (z,\bar z) = \lim_{\omega\rightarrow 0} \omega a^{out}(p, \sigma=+2) = \lim_{\omega\rightarrow 0+} \sqrt{2\pi^2} \ \int_{-\infty}^{\infty}d\lambda \ \omega^{i\lambda} e^{i\omega U} A^{out}_{\lambda,\sigma = +2}(U,z,\bar z)
\ee

Using this operator we can write Eq-$\ref{SG}$ as,
\be
\begin{gathered}
\bra{\Omega} O_{+}(z,\bar z) \prod_{i=M+1}^{M+N} A^{out}_{n_i,\lambda_i,\sigma_i}(U_i,z_i,\bar z_i) \prod_{j=1}^{M}A^{\dagger in}_{n_j,\lambda_j,\sigma_j}(U_j,z_j,\bar z_j) \ket{\Omega} \\
=\frac{1}{(\sqrt{8\pi^4})^{M+N}} \bigg(\prod_{i=M+1}^{M+N}\int_{0}^{\infty}d\omega_i \omega_i^{- i\lambda_i}e^{-i\omega_i U_i}\bigg)\bigg(\prod_{j=1}^{M}\int_{0}^{\infty}d\omega_j \omega_j^{i\lambda_j} e^{i\omega_j U_j}\bigg) \\
\bigg\{\frac{\kappa}{2}\bigg[ \sum_{j=1}^{M}\frac{\omega_j(\bar z- \bar z_j)}{z-z_j} - \sum_{i=M+1}^{M+N}\frac{\omega_i(\bar z- \bar z_i)}{z-z_i}\bigg] S_{N\leftarrow M}\big(\ p_i,\sigma_i,n_i\}, out \ \big| \ \{p_j,\sigma_j,n_j\}, in \big)\bigg\}
\end{gathered}
\ee

Now applying $\bar\partial_z$ to both sides we get,
\be
\begin{gathered}
\bra{\Omega}\bar\partial_z O_{+}(z,\bar z) \prod_{i=M+1}^{M+N} A^{out}_{n_i,\lambda_i,\sigma_i}(U_i,z_i,\bar z_i) \prod_{j=1}^{M}A^{\dagger in}_{n_j,\lambda_j,\sigma_j}(U_j,z_j,\bar z_j) \ket{\Omega} \\
=\frac{1}{(\sqrt{8\pi^4})^{M+N}} \bigg(\prod_{i=M+1}^{M+N}\int_{0}^{\infty}d\omega_i \omega_i^{- i\lambda_i}e^{-i\omega_i U_i}\bigg)\bigg(\prod_{j=1}^{M}\int_{0}^{\infty}d\omega_j \omega_j^{i\lambda_j} e^{i\omega_j U_j}\bigg)\\
\bigg\{\frac{\kappa}{2}\bigg[ \sum_{j=1}^{M}\frac{\omega_j}{z-z_j} - \sum_{i=M+1}^{M+N}\frac{\omega_i}{z-z_i}\bigg] S_{N\leftarrow M}\big(\{p_i,\sigma_i,n_i\}, out \ \big| \ \{p_j,\sigma_j,n_j\}, in \big)\bigg\}\\\\
= \frac{\kappa}{2}\bigg[ \sum_{j=1}^{M}\frac{-i \partial / \partial U_j}{z-z_j} + \sum_{i=M+1}^{M+N}\frac{-i\partial / \partial U_i}{z-z_i}\bigg] \\
\bigg[\frac{1}{(\sqrt{8\pi^4})^{M+N}} \bigg(\prod_{i=M+1}^{M+N}\int_{0}^{\infty}d\omega_i \omega_i^{- i\lambda_i}e^{-i\omega_i U_i}\bigg)\bigg(\prod_{j=1}^{M}\int_{0}^{\infty}d\omega_j \omega_j^{i\lambda_j} e^{i\omega_j U_j}\bigg)\\
 S_{N\leftarrow M}\big(\ p_i,\sigma_i,n_i\}, out \ \big| \ \{p_j,\sigma_j,n_j\}, in \big)\bigg]\\\\
= \frac{\kappa}{2}\bigg[ \sum_{j=1}^{M}\frac{-i \partial / \partial U_j}{z-z_j} + \sum_{i=M+1}^{M+N}\frac{-i\partial / \partial U_i}{z-z_i}\bigg] \bra{\Omega}\prod_{i=M+1}^{M+N} A^{out}_{n_i,\lambda_i,\sigma_i}(U_i,z_i,\bar z_i) \prod_{j=1}^{M}A^{\dagger in}_{n_j,\lambda_j,\sigma_j}(U_j,z_j,\bar z_j) \ket{\Omega}
\end{gathered} 
\ee

Therefore Weinberg's soft-graviton theorem takes the following form,

\be
\begin{gathered}
\bra{\Omega}\bar\partial O_{+}(z,\bar z) \prod_{i=M+1}^{M+N} A^{out}_{n_i,\lambda_i,\sigma_i}(U_i,z_i,\bar z_i) \prod_{j=1}^{M}A^{\dagger in}_{n_j,\lambda_j,\sigma_j}(U_j,z_j,\bar z_j) \ket{\Omega} \\
= \frac{\kappa}{2}\bigg[ \sum_{j=1}^{M}\frac{-i \partial / \partial U_j}{z-z_j} + \sum_{i=M+1}^{M+N}\frac{-i\partial / \partial U_i}{z-z_i}\bigg] \bra{\Omega}\prod_{i=M+1}^{M+N} A^{out}_{n_i,\lambda_i,\sigma_i}(U_i,z_i,\bar z_i) \prod_{j=1}^{M}A^{\dagger in}_{n_j,\lambda_j,\sigma_j}(U_j,z_j,\bar z_j) \ket{\Omega} 
\end{gathered}
\ee

%This relation means that the insertion of the operator $\bar\partial_z O_{+}(z,\bar z)$  

Now using the relation 
\be
\bar\partial_z \frac{1}{z-w} = 2\pi \delta^2(z-w)
\ee

it is easy to check that the insertion of the operator $T(g)$, given by

\be
T(g) = \frac{i}{\pi\kappa} \int d^2 z g(z,\bar z) \bar\partial^2 O_{+}(z,\bar z)
\ee

leads to

\be\label{ST}
\boxed{
\begin{gathered}
\bra{\Omega}T(g) \prod_{i=M+1}^{M+N} A^{out}_{n_i,\lambda_i,\sigma_i}(U_i,z_i,\bar z_i) \prod_{j=1}^{M}A^{\dagger in}_{n_j,\lambda_j,\sigma_j}(U_j,z_j,\bar z_j) \ket{\Omega} \\
= \bigg[ \sum_{j=1}^{M} g(z_j, \bar z_j)\frac{\partial}{\partial U_j} + \sum_{i=M+1}^{M+N}g(z_i,\bar z_i)\frac{\partial}{\partial U_i}\bigg] \bra{\Omega}\prod_{i=M+1}^{M+N} A^{out}_{n_i,\lambda_i,\sigma_i}(U_i,z_i,\bar z_i) \prod_{j=1}^{M}A^{\dagger in}_{n_j,\lambda_j,\sigma_j}(U_j,z_j,\bar z_j) \ket{\Omega} 
\end{gathered}}
\ee 

This relation means that the insertion of the operator $T(g)$ generates a point transformation in the $(U,z,\bar z)$ space, given by, 
\be
\big(U,z,\bar z\big) \rightarrow \big(U + g(z,\bar z), z, \bar z \big)
\ee

where $g(z,\bar z)$ is an \textit{arbitrary} function on the plane. If we think of the $(U,z,\bar z)$ space as the null-infinity in Minkowski space then these transformations can be identified with BMS supertranslations \cite{Strominger:2013jfa,Bondi:1962px,Barnich:2009se}.

Now as Eq-$\ref{ST}$ shows the $\tilde A$-amplitudes are not generically invariant under this transformation \cite{Strominger:2013jfa}. But there are four transformations, corresponding to four global space-time translations, under which $\tilde A$ remains unchanged. To see this let us first note that in terms of the variable, $U = u(1+z\bar z)$, global space-time translation by a 4-vector $l^{\mu}$ is given by, 
\be
U \rightarrow U + (l^0 - l^3) - (l^1 - i l^2) z - (l^1 + i l^2) \bar z + (l^0 + l^3) z \bar z , \ z\rightarrow z, \ \bar z\rightarrow \bar z
\ee
Now let $g(z,\bar z,l)$ denote the function $g(z,\bar z)$ corresponding to the global space-time translation by 4-vector $l^{\mu}$. Then, 
\be
g(z,\bar z,l) = (l^0 - l^3) - (l^1 - i l^2) z - (l^1 + i l^2) \bar z + (l^0 + l^3) z \bar z
\ee
It is also easy to see that, 
\be
{\bar\partial}^2 g(z,\bar z,l) = {\partial}^2 g(z,\bar z,l) = 0
\ee

Using $g(z,\bar z,l)$ we can write the identity given in Eq-$\ref{ST}$ as, 

\be
\begin{gathered}
\frac{i}{\pi\kappa} \int d^2 z \ g(z,\bar z,l) \bar\partial^2\bra{\Omega} O_{+}(z,\bar z)\prod_{i=M+1}^{M+N} A^{out}_{n_i,\lambda_i,\sigma_i}(U_i,z_i,\bar z_i) \prod_{j=1}^{M}A^{\dagger in}_{n_j,\lambda_j,\sigma_j}(U_j,z_j,\bar z_j) \ket{\Omega} \\
= \bigg[ \sum_{j=1}^{M} g(z_j, \bar z_j,l)\frac{\partial}{\partial U_j} + \sum_{i=M+1}^{M+N}g(z_i,\bar z_i,l)\frac{\partial}{\partial U_i}\bigg] \bra{\Omega}\prod_{i=M+1}^{M+N} A^{out}_{n_i,\lambda_i,\sigma_i}(U_i,z_i,\bar z_i) \prod_{j=1}^{M}A^{\dagger in}_{n_j,\lambda_j,\sigma_j}(U_j,z_j,\bar z_j) \ket{\Omega} 
\end{gathered}
\ee
 
Now assuming that we are allowed to integrate by parts we get,

\be
\begin{gathered}
\frac{i}{\pi\kappa} \int d^2 z \bar\partial^2 g(z,\bar z,l) \bra{\Omega} O_{+}(z,\bar z)\prod_{i=M+1}^{M+N} A^{out}_{n_i,\lambda_i,\sigma_i}(U_i,z_i,\bar z_i) \prod_{j=1}^{M}A^{\dagger in}_{n_j,\lambda_j,\sigma_j}(U_j,z_j,\bar z_j) \ket{\Omega} \\
= \bigg[ \sum_{j=1}^{M} g(z_j, \bar z_j,l)\frac{\partial}{\partial U_j} + \sum_{i=M+1}^{M+N}g(z_i,\bar z_i,l)\frac{\partial}{\partial U_i}\bigg] \bra{\Omega}\prod_{i=M+1}^{M+N} A^{out}_{n_i,\lambda_i,\sigma_i}(U_i,z_i,\bar z_i) \prod_{j=1}^{M}A^{\dagger in}_{n_j,\lambda_j,\sigma_j}(U_j,z_j,\bar z_j) \ket{\Omega} 
\end{gathered}
\ee

Now using $\bar\partial^2 g(z,\bar z,l)=0$ we get, 

\be
\boxed{
\bigg[ \sum_{j=1}^{M} g(z_j, \bar z_j,l)\frac{\partial}{\partial U_j} + \sum_{i=M+1}^{M+N}g(z_i,\bar z_i,l)\frac{\partial}{\partial U_i}\bigg] \bra{\Omega}\prod_{i=M+1}^{M+N} A^{out}_{n_i,\lambda_i,\sigma_i}(U_i,z_i,\bar z_i) \prod_{j=1}^{M}A^{\dagger in}_{n_j,\lambda_j,\sigma_j}(U_j,z_j,\bar z_j) \ket{\Omega}  = 0}
\ee

This is essentially the statement of the invariance of the $\tilde A$ amplitude (or the S-matrix in the Fock basis) under global space-time translations. 

\section{A slightly different treatment of the soft-operator}

Let us again consider the equation,
\be\label{SP}
\begin{gathered}
\lim_{\omega\rightarrow 0+}\sqrt{2\pi^2} \int_{-\infty}^{\infty}d\lambda \ \omega^{i\lambda} e^{i\omega U} \bra{\Omega} A^{out}_{\lambda,\sigma}(U,z,\bar z) \prod_{i=M+1}^{M+N} A^{out}_{n_i,\lambda_i,\sigma_i}(U_i,z_i,\bar z_i) \prod_{j=1}^{M}A^{\dagger in}_{n_j,\lambda_j,\sigma_j}(U_j,z_j,\bar z_j) \ket{\Omega} \\
=\frac{1}{(\sqrt{8\pi^4})^{M+N}} \bigg(\prod_{i=M+1}^{M+N}\int_{0}^{\infty}d\omega_i \omega_i^{- i\lambda_i}e^{-i\omega_i U_i}\bigg)\bigg(\prod_{j=1}^{M}\int_{0}^{\infty}d\omega_j \omega_j^{i\lambda_j} e^{i\omega_j U_j}\bigg) \\
\lim_{\omega\rightarrow 0+} \omega S_{N+1\leftarrow M}\big(\{p,\sigma \ ; \ p_i,\sigma_i,n_i\}, out \ \big| \ \{p_j,\sigma_j,n_j\}, in \big)
\end{gathered}
\ee
and study the following operator. 
\be
J_{\sigma, \omega}(U,z,\bar z) = \sqrt{2\pi^2} \ \int_{-\infty}^{\infty}d\lambda \ \omega^{i\lambda} e^{i\omega U}A^{out}_{\lambda,\sigma}(U,z,\bar z) 
\ee
This is the same as $\omega a^{out}(p,\sigma)$ but we do not want to use this fact in this section. This is the reason why we have kept the $U$ dependence in $J_{\sigma,\omega}$. The only thing that we want to use is the Lorentz transformation property of the objects involved in Eq-$\ref{SP}$. We will show, using the Lorentz transformation property of $A_{\lambda,\sigma}$, that in the soft limit ($\omega\rightarrow 0+$) the soft operator defined as $J_{\sigma}(U,z,\bar z) = J_{\sigma,\omega=0}(U,z,\bar z)$ transforms as a tensor in the $(U,z,\bar z)$ space. Then we invoke the soft theorem to show that the tensor $J_{\sigma}(U,z,\bar z)$ does not depend on $U$, i.e, $\partial_{U}J_{\sigma}(U,z,\bar z)=0$. This is an interesting phenomenon because no such thing can happen in a Lorentz invariant theory defined on Minkowski space. The condition that a non-trivial tensor function on Minkowski space be independent of time is not Lorentz invariant. But the equation $\partial_{U}J_{\sigma}(U,z,\bar z)=0$ is in fact Lorentz invariant in $(U,z,\bar z)$ space. So the following demonstration can be thought of as an overall consistency check of the procedure. \\

Using the transformation property of $A_{\lambda,\sigma}$ we get, 
\be
U(\Lambda) J_{\sigma,\omega}(U,z,\bar z) U(\Lambda)^{-1} = \frac{1}{(cz+d)^{2h}}\frac{1}{(\bar c \bar z + \bar d)^{2\bar h}} J_{\sigma, \omega |cz+d|^2}\bigg(\frac{U}{|cz+d|^2},\frac{az+b}{cz+d}, \ \frac{\bar a \bar z + \bar b}{\bar c \bar z + \bar d}\bigg)
\ee
where $U(\Lambda)$ is a $SL(2,\mathbb{C})$ transformation and
\be
h = \frac{1+\sigma}{2}, \ \bar h = \frac{1-\sigma}{2}
\ee

Now let us take the soft limit $\omega\rightarrow 0+$ and define the soft operator $J_{\sigma}(U,z,\bar z)$ to be $J_{\sigma, \omega =0}(U,z,\bar z)$. Under $SL(2,\mathbb{C})$ transformation, 
\be
U(\Lambda) J_{\sigma}(U,z,\bar z) U(\Lambda)^{-1} = \frac{1}{(cz+d)^{2h}}\frac{1}{(\bar c \bar z + \bar d)^{2\bar h}} J_{\sigma}\bigg(\frac{U}{|cz+d|^2},\frac{az+b}{cz+d}, \ \frac{\bar a \bar z + \bar b}{\bar c \bar z + \bar d}\bigg)
\ee

We can also study the transformation of $J_{\sigma,\omega}(U,z,\bar z)$ under global translation by a four vector $l^{\mu}$. The transformation property is given by,
\be
e^{-il.P} J_{\sigma,\omega} (U,z,\bar z) e^{il.P} = e^{-i\omega g(z,\bar z, l)} J_{\sigma,\omega} (U + g(z,\bar z,l),z,\bar z)
\ee
where
\be
g(z,\bar z,l) = (l^0 - l^3) - (l^1 - i l^2) z - (l^1 + i l^2) \bar z + (l^0 + l^3) z \bar z
\ee
So in the soft limit $\omega\rightarrow 0+$ we get,
\be
e^{-il.P} J_{\sigma} (U,z,\bar z) e^{il.P} = J_{\sigma} (U + g(z,\bar z,l),z,\bar z)
\ee

So we can see that the soft-operator $J_{\sigma}(U,z,\bar z)$ transforms like an $ISL(2,\mathbb{C})$ primary as defined in \cite{Banerjee:2018gce}.

Now soft theorem tells us that the correlator 
\be
\bra{\Omega} J_{\sigma}(U,z,\bar z) \prod_{i=M+1}^{M+N} A^{out}_{n_i,\lambda_i,\sigma_i}(U_i,z_i,\bar z_i) \prod_{j=1}^{M}A^{\dagger in}_{n_j,\lambda_j,\sigma_j}(U_j,z_j,\bar z_j) \ket{\Omega}
\ee

does \textit{not} depend on $U$. So we can write $\partial_{U}J_{\sigma}(U,z, \bar z) = 0$ as an operator equation because its matrix elements between arbitrary in and out states is zero. Let us now show that the condition $\partial_{U}J_{\sigma}(U,z, \bar z) = 0$ is in fact Lorentz invariant.

We can prove this more generally for any tensor $\phi_{h,\bar h}(U,z,\bar z)$ which satisfies $\partial_{U}\phi_{h, \bar h}(U,z,\bar z)=0$.

Under $SL(2,\mathbb{C})$ transformation, 
\be
\phi'_{h,\bar h} (U',z',\bar z') \bigg(\frac{dz'}{dz}\bigg)^{h} \bigg(\frac{d\bar z'}{d\bar z}\bigg)^{\bar h} = \phi_{h,\bar h} (U,z,\bar z)
\ee
where 
\be
(U',z',\bar z') = \bigg(\frac{U}{|cz+d|^2} , \frac{az+b}{cz+d} , \frac{\bar a \bar z + \bar b}{\bar c \bar z + \bar d}\bigg)
\ee
%Here we have taken $(h,\bar h)$ to be any two complex numbers subject to the restriction that $h-\bar h \in \mathbb{Z}/2$.

 Therefore, 
  \be
 \phi'_{h,\bar h} (U',z',\bar z') = (cz+d)^{2h} (\bar c \bar z + \bar d)^{2\bar h} \phi_{h,\bar h} (U,z,\bar z)
 \ee
 and 
 \be
 \frac{\partial}{\partial U'} = \frac{\partial U}{\partial U'} \frac{\partial}{\partial U} +\frac{\partial z}{\partial U'} \frac{\partial}{\partial z} + \frac{\partial \bar z}{\partial U'} \frac{\partial}{\partial \bar z}  = |cz+d|^2 \frac{\partial}{\partial U}
 \ee
 Using this we get, 
\be
\partial_{U'} \phi'_{h,\bar h} (U',z',\bar z') = |cz+d|^2 (cz+d)^{2h} (\bar c \bar z + \bar d)^{2\bar h} \ \partial_{U} \phi_{h,\bar h}(U,z,\bar z) 
\ee
 
 So $\partial_U\phi_{h,\bar h}$ transforms homogeneously under Lorentz transformation. So in the $(U,z,\bar z)$ space, if it is zero in one frame then it is zero in any other frame. So $J_{\sigma}(U,z,\bar z) = J_{\sigma}(z,\bar z)$.
 
 This shows that in the $(U,z,\bar z)$ space we have well-defined Lorentz tensors of the form $\phi_{h,\bar h}(z,\bar z)$.
 
\section{Discussion} 
In this paper we have tried to connect the lading soft theorems to symmetries in the approach suggested by \cite{Pasterski:2016qvg,Pasterski:2017kqt,Pasterski:2017ylz,Banerjee:2018gce}. This does This seems to raise more questions than it answers. Let us list some of the pressing questions : 

%1) It seems that the derivation in this paper is somewhat heuristic, but the end results are the expected ones. It will be nice to make it as rigorous as possible. 

1) We have not dealt with the subleading soft theorems. There are fascinating new results \cite{Cachazo:2014fwa,Schwab:2014xua,Bern:2014oka,Broedel:2014fsa,Sen:2017xjn,Sen:2017nim,Laddha:2017ygw,Chakrabarti:2017ltl,Chakrabarti:2017zmh,Laddha:2017vfh,Laddha:2018rle,Campoleoni:2017mbt} in this subject. In particular it will be interesting to understand subleading soft graviton theorem and its relation to superrotation \cite{Barnich:2009se,Banks:2003vp,deBoer:2003vf,Cheung:2016iub,Kapec:2016jld,Kapec:2014opa,Kapec:2017gsg,Pasterski:2015tva,Campiglia:2014yka} from this point of view. 

2) The results of \cite{Pasterski:2016qvg,Pasterski:2017kqt,Pasterski:2017ylz} are also applicable to the massive particles. It will be interesting to extend the results of \cite{Banerjee:2018gce} to the massive case \cite{SB}.

3) We have worked with creation and annihilation free fields on the $(u,z,\bar z)$ space. In quantum field theory on Minkowski space we usually work with quantum fields built out of both creation and annihilation operators. This is required by causality and locality in space-time. 

Can one define a reasonable notion of causality and locality in the $(u,z,\bar z)$ space ? This should be possible because the Poincare group acts on this space. Using this, can one define local quantum (free) fields on this space ? To what extent the interacting theory make sense ? If it makes sense then its (global) symmetry structure should be closely related to the (global) asymptotic symmetries of quantum gravity in flat space. 

4) It will be interesting to better understand the unitary representation of the four dimensional BMS group including superrotations \cite{Barnich:2014kra,Campoleoni:2016vsh,Bagchi:2016geg}.

5) Is there any possible holographic interpretation \cite{Pasterski:2016qvg,Cheung:2016iub,deBoer:2003vf,Kapec:2017gsg,Bagchi:2016bcd} ?

%\section{Comments On Locality For Bosonic Fields} 

%\be
%\phi_{\lambda,\sigma}(U,z,\bar z) = A^{\dagger}_{\lambda,\sigma}(U,z,\bar z) + A_{-\lambda,-\sigma}(U,z,\bar z)
%\ee

%\be
%\phi^{\dagger}_{\lambda,\sigma}(U,z,\bar z) = \phi_{-\lambda,-\sigma}(U,z,\bar z)
%\ee
 
%\be
%U(\Lambda) \ \phi_{\lambda,\sigma}(U, z,\bar z) \ U(\Lambda)^{-1} \\
% = \frac{1}{(cz+d)^{2h}} \frac{1}{(\bar c \bar z + \bar d)^{2 \bar h}}  \ \phi_{\lambda,\sigma}\bigg(\frac{U}{|cz+d|^2}, \frac{az+b}{cz+d} \ , \frac{\bar a \bar z + \bar b}{\bar c \bar z + \bar d}\bigg)
%\ee 

% \be
%h = \frac{1+ i\lambda - \sigma}{2} , \  \bar h = \frac{1+ i\lambda + \sigma}{2}
%\ee
 
% \be
%\begin{gathered}
% [\phi_{\lambda,\sigma} (U, z, \bar z), \phi_{\lambda',\sigma'}(U', z', \bar z')] 
%= \frac{\delta_{\sigma + \sigma'}}{2\pi} \Gamma\big(i(\lambda' + \lambda)\big)\delta^2(z-z') \Delta_{\lambda,\lambda'}(U'-U)
%\end{gathered}
% \ee
 
%\be
%\Delta_{\lambda,\lambda'}(U'-U) = \frac{1}{\big(-i(U'-U + i\epsilon)\big)^{i(\lambda' + \lambda)}} - \frac{1}{\big(-i(U-U' + i\epsilon)\big)^{i(\lambda' + \lambda)}} = - \Delta_{\lambda,\lambda'}(U-U')
%ee

\section{Acknowledgement}
I would like to thank especially Ashoke Sen for numerous enlightening discussions on soft theorems and related subjects. I would also like to thank Alok Laddha and Prahar Mitra for very helpful discussion on soft theorems and related matters. Part of this work was presented in the "First Spring Meeting on Strings" held in NISER Bhubaneswar, India. I would like to thank the organizers for holding this exciting meeting. I would also like to thank all the participants of the meeting especially Jyotirmoy Bhattacharya, Sayantani Bhattacharya, Bobby Ezhuthachan, Arnab Kundu, Jnan Maharana, Sudhakar Panda, Ashoke Sen and Yogesh Srivastava for their valuable feedback.

\end{document}